\documentstyle[12pt]{article}
\newcommand\be{\begin{equation}}
\newcommand\ee{\end{equation}}
\newcommand\bea{\begin{eqnarray}}
\newcommand\eea{\end{eqnarray}}
\newcommand\ket[1]{|#1\rangle}
\newcommand\bra[1]{\langle #1|}

\newcommand{\fatalpha}{{\bf \alpha \kern -0.44em \alpha}}
\newcommand{\fatsigma}{{\bf \sigma \kern -0.54em \sigma}}
\newcommand{\tpchi}{{\bf \chi \kern -0.35em \chi}}
\newcommand{\llambda}{{\bf \lambda \kern -0.45em \lambda}}

\renewcommand{\theequation}{\arabic{equation}}
\renewcommand{\theequation}{\thesection-\arabic{equation}}
\bibliography{plain}
\title{\bf Bell states diagonal entanglement
witnesses } \vspace{20mm}
\author{ M. A. Jafarizadeh$^{a,b,c}$
 \thanks{E-mail:jafarizadeh@tabrizu.ac.ir},
 M. Rezaee$^{a,b}$ \thanks{E-mail:karamaty@tabrizu.ac.ir}, S. K. A. Seyed Yagoobi$^{a}$ \thanks{E-mail:skasy@tabrizu.ac.ir}
\\
\\
$^a${\small Department of Theoretical Physics and Astrophysics,
Tabriz University, Tabriz 51664, Iran.} \\ $^b${\small Institute
for Studies in Theoretical Physics and Mathematics, Tehran
19395-1795, Iran.} \\ $^c${\small Research Institute for
Fundamental Sciences, Tabriz 51664, Iran.}} \pagebreak

\begin{document}
\maketitle \vspace{15mm}
\newpage
\begin{abstract}

It has been shown that finding  generic  Bell states diagonal
entanglement witnesses (BDEW) for $d_{1}\otimes d_{2}\otimes
....\otimes d_{n}$ systems  exactly reduces to a linear
programming if the feasible region be a polygon by itself and
approximately obtains via linear programming if the feasible
region is not a polygon. Since solving linear programming for
generic case is difficult, the multi-qubits, $2\otimes N$ and $3
\otimes 3$ systems for the special case of generic BDEW for some
particular choice of their parameters have been considered. In
the rest of this paper we obtain the optimal non decomposable
entanglement witness for $3 \otimes 3$ system for some particular
choice of its parameters. By proving the optimality of the well
known reduction map and combining it with the  optimal and
non-decomposable 3 $\otimes$ 3 BDEW (named critical entanglement
witnesses)  the family of optimal and non-decomposable 3
$\otimes$ 3 BDEW have also been obtained. Using the approximately
critical entanglement witnesses, some 3 $\otimes$ 3 bound
entangled states are so detected. So the well known Choi map  as
a particular case of the positive map in connection with this
witness via Jamiolkowski isomorphism has been considered which
approximately is obtained via linear programming.

 {\bf Keywords: Entanglement witness, Bell
decomposable state, non decomposable entanglement witness,
Optimal entanglement witness, Choi map.}

{\bf PACs Index: 03.65.Ud }
\end{abstract}
\newpage
\vspace{70mm}
\section{Introduction}
Entanglement is one of the most fascinating features of quantum
mechanics. As Einstein, Podolsky and Rosen \cite{Einstein}
pointed out, the quantum states of two physically separated
systems that interacted in the past can defy our intuitions about
the outcome of local measurements. Moreover, it has recently been
recognized that entanglement is a very important resource in
quantum information processing\cite{nielsen}. A bipartite mixed
state is said to be separable \cite{werner} (not entangled) if it
can be written as a convex combination of pure product states.

A separability criterion is based on a simple property that can
be shown to hold for every separable state. If some state does
not satisfy this property, then it must be entangled. But the
converse does not necessarily imply the state to be separable. One
of the first and most widely used related criterion is the
Positive Partial Transpose (PPT) criterion, introduced by Peres
\cite{peres}. Furthermore, the necessary and sufficient condition
for separability in $H_2 \otimes H_2$ and $H_2 \otimes H_3$ was
shown by  Horodeckis \cite{horodecki1}, which was based on a
previous work by Woronowicz \cite{woronowicz}. However, in higher
dimensions, there are PPT states that are nonetheless entangled,
as was first shown in \cite{horodecki2},  based on
\cite{woronowicz}. These states are called bound entangled states
because they have the peculiar property that no entanglement can
be distilled from them by local operations \cite{horodecki3}.

Another approach to distinguish separable states from entangled
states involves the so called entanglement witness (EW)
\cite{terhal}. An EW  for a given entangled state $\rho$ is an
observable W whose expectation value is nonnegative on any
separable state, but strictly negative on an entangled state
$\rho$.

There is a correspondence  relating entanglement witnesses to
linear positive (but not completely positive) maps from the
operators on $H_A$ to the operators on $H_B$ via  Jamiolkowski
isomorphism, or vice versa\cite{jamiolkowski}.

There has been much work on the separability problem,
particularly from the Innsbruck-Hannover group, as reviewed in
\cite{terhal2,bruss}, that emphasizes convexity and proceeds by
characterizing entanglement witnesses in terms of their extreme
points, the so-called optimal entanglement
witnesses\cite{Doherty}, and PPT entangled states in terms of
their extreme points, the edge PPT entangled states
\cite{cirac,kraus}.

Having constructed the EW , one can decompose it into a sum of
local measurements, then the expectation value can be measured
with simple method. This decomposition has to be optimized in a
certain way since we want to use the smallest number of
measurements possible\cite{Otfrid1,Otfrid2,Otfrid3,Hyllus}.

In this paper, we show that finding  generic  Bell states diagonal
entanglement witnesses (BDEW) for $d_{1}\otimes d_{2}\otimes
....\otimes d_{n}$ systems reduces to a convex optimization
problem. If the feasible region for this optimization problem
constructs a polygon by itself, the corresponding boundary points
of the convex hull will minimize exactly  optimization problem.
This problem is called linear programming , and the simplex
method is the easiest way of solving it. If the feasible region
is not a polygon, with the help of tangent planes in this region
at points which are determined either analytically or numerically
one can define new convex hull which is a polygon and has
encircled  the feasible region. The points on the boundary of the
polygon can approximately determine the minimum value of
optimization problem. Thus approximated value is obtained via
linear programming. In general it is difficult to find this
region and solve the optimization problem, thus it is difficult
to find any generic multipartite EW. In the following sections we
consider some simple but important examples which are solved
exactly or approximately by linear programming method. Then we
consider the multi-qubits and  $2\otimes N$ with exactly minimum
value by linear programming  and $3 \otimes 3$ systems  with
approximately minimum value by linear programming and then
establish $3\otimes 3$ optimality condition together with
non-decomposability properties  for some particular choice of its
parameters. Then we combine  the optimal  well known reduction
map, and the  optimal as well as the non-decomposable 3 $\otimes$
3 BDEW (i.e., the critical entanglement witnesses) to obtain
further family of optimal and non-decomposable 3 $\otimes$ 3
BDEW. Finally, using the critical entanglement witnesses some 3
$\otimes$ 3 bound entangled states are detected and  we consider
the well known Choi map  as a particular case of the positive map
in connection with this witness via Jamiolkowski isomorphism
which  approximately is obtained via linear programming.

 The paper is organized as follows:\\ In
section 2 we give a brief review of  entanglement witness. In
section 3 we show that finding  generic  Bell states diagonal
entanglement witnesses for $d_{1}\otimes d_{2}\otimes ....\otimes
d_{n}$ systems reduces to a linear programming problem.  In
section 4, we consider BDEW for multi-qubit system. In section 5,
we provide BDEW for $2\otimes N$. In section 6, we provide BDEW
for $3\otimes 3$ systems. Section 7 is devoted to prove the n-d
of critical EW and introduce a new family of optimal nd-EW via
combining critical EW with the well known reduction maps. In
section 8, using the critical EW, we will be able to detect a
bound BD entangled state. In section 9, we consider the well
known Choi map  as a particular case of the positive map connect
with this witness via Jamiolkowski isomorphism.  Finally in
section 10 using the optimal EW, we show that some separable Bell
states diagonal lies at the boundary of separable region. The
paper is ended with a brief conclusion together with three
appendices devoted to the proof of  A) the optimization of product
distributions B)optimality of critical, reduction map C)simplex
method for solving linear programming problem.
\section{Entanglement witness}
Here we  mention briefly those concepts and definitions of EW that
will be needed in the sequel, a more detailed treatment may be
found for example in \cite{woronowicz,jamiolkowski,lec}.

Let S be a convex compact set in a finite dimensional Banach
space. Let $\rho$ be a point in the space with $\rho\;\;\mbox
{which is not in}\;\;S$. Then there exists a hyperplane\cite{lec}
that separates $\rho$ from S.

A hermitian operator (an observable) W is called an entanglement
witness (EW) iff \be   \exists\rho\;\mbox{such that}\;\;
Tr(\hat{\rho}{W} ) < 0\ee \be \forall {\rho^{\prime}} \in
S\;\;\;Tr({\rho^{\prime}}\hat{W} )\geq 0.\ee

{\bf{Definition 1:}} An EW is decomposable iff there exists
operators P, Q such that \be W=P+Q^{T_{A}}\;\;\;\;\;P,Q>0. \ee
Decomposable EW can not detect PPT entangled
states\cite{woronowicz}.

{\bf{Definition 2:}} An EW is called non-decomposable entanglement
witness (nd-EW) iff there exists at least one PPT entangled state
which the witness detects\cite{woronowicz}.

{\bf{Definition 3:}} The (decomposable) entanglement witness is
tangent to S (P) iff there exists a $\sigma\in S$ ( $\rho\in P$)
with $Tr (W\sigma) = 0 \;\;(Tr (W\rho) = 0)$.

Using these definitions we can restate the consequences of the
Hahn-Banach theorem \cite{lec} in several ways:

{\bf{Theorem:}}

1- $\rho$ is entangled iff there exists a witness W such that $Tr
(\rho W) < 0$.

2- $\rho$ is a PPT entangled state iff there exists  an
non-decomposable entanglement witness W such that $Tr (\rho W) <
0$.

3- $\sigma$ is separable iff for all EW $\;\;Tr (W\sigma) \geq 0$.

From  theoretical point of view this theorem is quite powerful.
However, it is not useful to construct witnesses that detect a
given state $\rho$.

We know that  a strong relation was developed between entanglement
witnesses and positive maps\cite{woronowicz,jamiolkowski}. Notice
that an entanglement witness only gives one condition (namely $Tr
(W\rho) < 0$) while for the  map $(I_{A} \otimes \phi)\rho$ to be
positive definite, there are many conditions that have to  be
satisfied. Thus the map is much stronger, while the witnesses are
much weaker in detecting entanglement.  It is shown that this
concept is able to provide a more detailed classification of
entangled states.
\section{Bell states diagonal entanglement
witnesses} As we know, one can expand any trace class observable
in the Bell basis as  \be W=\sum_{_{i_{1}i_{2}...i_{n}}}
W_{_{i_{1}i_{2}...i_{n}}}
\ket{\psi_{_{i_{1}i_{2}...i_{n}}}}\bra{\psi_{_{i_{1}i_{2}...i_{n}}}}
\ee where $\ket{\psi_{_{i_{1}i_{2}...i_{n}}}}$(${0\leq i_{1}\leq
d_{1}, 0\leq i_{2}\leq d_{2},...,0\leq i_{n}\leq d_{n},}$ and
$d_{1}\leq d_{2}\leq ... \leq d_{n}$) stands for the orthonormal
states for a $d_{1}\otimes d_{2} ...\otimes d_{n}$ Bell state
defined as \be
\ket{\psi_{_{i_{1}i_{2}...i_{n}}}}=(\Omega)^{i_{1}}\otimes
(S)^{i_{2}}\otimes ... \otimes
(S)^{i_{n}}\ket{\psi_{_{00...0}}}\ee where $\Omega$ and S are
phase modules and shift operators for a $d_{1}\otimes d_{2}\otimes
....\otimes d_{n}$ defined as \be
\begin{array}{cc}\Omega=\left( \begin{array}{ccccc} 1 & 0 & 0 & \ldots & 0 \\  0 & \omega
& 0 & \ldots & 0\\ \vdots & \vdots  & \vdots & \ddots & \vdots\\ 0
& 0 & 0 &  \ldots & \omega^{d-1}\end{array}\right), & S=\left(
\begin{array}{ccccc} 0 & 1 & 0 & \ldots & 0 \\  0 & 0 & 1 & \ldots & 0\\ \vdots & \vdots & \vdots & \ddots & \vdots\\ 1 & 0 &
0 & \ldots & 0\end{array}\right),\end{array} \ee with
$\omega=exp(\frac{2\pi i}{d})$ and \be
\ket{\psi_{_{00...0}}}=\frac{1}{\sqrt{d}}\sum_{
_{i=0}}^{d_{1}-1}\ket{i}_{1}\ket{i}_{2}...\ket{i}_{n}.\ee W is a
trace one observable i.e., $Tr(W)=1$  and we have
$\sum_{_{i_{1}i_{2}...i_{n}}} W_{_{i_{1}i_{2}...i_{n}}} =1$.

Let us split the observable W into its positive and negative
spectra as: \be\label{11} W=\sum_{k=1}^{n^{+}}
\lambda_{k}^{+}\ket{\phi^{+}_{k}}\bra{\phi^{+}_{k}}-\sum_{k=1}^{n^{-}}
\mid\lambda_{k}^{-}\mid\ket{\phi^{-}_{k}}\bra{\phi^{-}_{k}},  \ee
where $\lambda_{k}^{+}(\lambda_{k}^{-})$ are the positive
(negative) eigenvalues $\ket{\phi^{+}_{k}}(\ket{\phi^{-}_{k}})$,
and we have $n^{+}+n^{-}=d^{n}$. Denoting $\sum\mid
\lambda^{-}_{k}\mid=s>0$ we  can  write (\ref{11}) as: \be
\label{wit4} W=(1+s)\rho^{+}-s\rho^{-}, \ee where $\rho^{\pm}$
are two normalized positive operators or density matrices defined
as \be
\rho^{+}=\frac{1}{1+s}\sum_{k=1}^{n^{+}}(\lambda^{+}_{k}\mid\phi^{+}_{k}><\phi^{+}_{k}\mid)\;\;,\;\;\rho^{-}=\frac{1}{s}\sum_{k=1}^{n^{-}}(\lambda^{-}_{k}\mid\phi^{-}_{k}><\phi^{-}_{k}\mid).
\ee Now, using the Lewenstein-Sanpera technique
\cite{lsd,lsd1,lsd2} the identity operator
$\frac{I_{d_{1}d_{2}...d_{n}}}{d_{1}d_{2}...d_{n}}$ can be
written in terms of $\rho^{-}$ and the other positive states as
 \be\label{wit5}
\frac{I_{d_{1}d_{2}...d_{n}}}{d_{1}d_{2}...d_{n}}=\lambda
\rho^{-}+(1-\lambda) \rho^{{\prime}^{-}}\;\;,\;\; 0<\lambda<1.
\ee By using the above equation we can replace $\rho^{-}$ in
Eq.(\ref{wit4}) in terms of the identity operator. So, Eq.(3-9)
is written as a sum of the identity and positive operators. Thus
we have \be\label{witt7}
W={\bf{r}}\frac{I_{d_{1}d_{2}...d_{n}}}{d_{1}d_{2}...d_{n}}+(1-{\bf{r}})\rho,\ee
where \be
\rho=\frac{(1+s)\lambda}{\lambda+s}\rho^{+}+s(\frac{1-\lambda}{s+\lambda})\rho^{{-}^{\prime}},\ee
and ${\bf{r}}=-\frac{s}{\lambda}<0$.

In this paper we have considered only trace one observables which
are diagonal in the Bell states. Hence we restrict ourselves to
the Bell states  diagonal $\rho$ defined as
\be\label{witt6}\rho=\sum_{_{i_{1}i_{2}...i_{n}}}
q_{_{i_{1}i_{2}...i_{n}}}\ket{\psi_{_{i_{1}i_{2}...i_{n}}}}\bra{\psi_{_{i_{1}i_{2}...i_{n}}}}\;\;,\;\;q_{_{i_{1}i_{2}...i_{n}}}>0\;\;
\mbox{and}\;\;\sum_{_{i_{1}i_{2}...i_{n}}}
q_{_{i_{1}i_{2}...i_{n}}}=1.\ee Finally, by substituting
(\ref{witt6}) in (\ref{witt7}) the trace one Bell states diagonal
 W observables are \be\label{GWE1}
W={\bf{r}}\frac{I_{d_{1}d_{2}...d_{n}}}{d_{1}d_{2}...d_{n}}+(1-{\bf{r}})\sum_{_{i_{1}i_{2}...i_{n}}}
q_{_{i_{1}i_{2}...i_{n}}}\ket{\psi_{_{i_{1}i_{2}...i_{n}}}}\bra{\psi_{_{i_{1}i_{2}...i_{n}}}}.
\ee The observable given by (\ref{GWE1}) is not a positive
operator and can not be an EW provided that its expectation value
on any pure product state is positive.
 For a given product state $\ket{\gamma}=\ket{\alpha}_{1}\ket{\alpha}_{2}...\ket{\alpha}_{n}$ the non negativity of  \be\label{optt1}
Tr(W\ket{\gamma}\bra{\gamma})\geq 0\ee implies that \be\label{p4}
\frac{-d_{1}d_{2}...d_{n}\sum_{_{i_{1}i_{2}...i_{n}}}q_{_{i_{1}i_{2}...i_{n}}}P_{_{i_{1}i_{2}...i_{n}}}}{1-d_{1}d_{2}...d_{n}\sum_{_{i_{1}i_{2}...i_{n}}}q_{_{i_{1}i_{2}...i_{n}}}P_{_{i_{1}i_{2}...i_{n}}}}
\leq {\bf{r}}\leq 0, \ee where
$P_{_{i_{1}i_{2}...i_{n}}}=\mid<\gamma\mid\psi_{_{i_{1}i_{2}...i_{n}}}>\mid^{2}$.

Denoting the summation in the  numerator  and the dominator in
(\ref{p4}) by
$\;\;\;\;\;\;\;\;C(\gamma)=d_{1}d_{2}...d_{n}\sum_{_{i_{1}i_{2}...i_{n}}}q_{_{i_{1}i_{2}...i_{n}}}P_{_{i_{1}i_{2}...i_{n}}}$
we see that the  least possible
${\bf{r}}_{0}=-\frac{C(\gamma)}{1-C(\gamma)}$ is the decreasing
function of $C(\gamma)$  for $C(\gamma) <1$ (obviously for
$C(\gamma) >1$ all {\bf{r}} while being  positive provide
positive expectation value). Therefore, for  given parameters
$\;\;q_{_{i_{1}i_{2}...i_{n}}}>0\;\;$,with
$\;\;\sum_{_{i_{1}i_{2}...i_{n}}}q_{_{i_{1}i_{2}...i_{n}}}=1\;\;\;\;$,
the least allowed value of the parameter ${\bf{r}}$, called the
critical parameter (denoted by ${\bf{r}}_{c}$ ) is obtained from
the product state $\gamma$ which minimizes
$C_{\gamma}=\sum_{_{i_{1}i_{2}...i_{n}}}q_{_{i_{1}i_{2}...i_{n}}}P_{_{i_{1}i_{2}...i_{n}}}$,
with $0\leq P_{_{i_{1}i_{2}...i_{n}}} \leq 1$ and the constraint
$\sum_{_{i_{1}i_{2}...i_{n}}}P_{_{i_{1}i_{2}...i_{n}}}=1$. As for
the completeness of the Bell state
$\sum_{_{i_{1}i_{2}...i_{n}}}\ket{\psi_{_{i_{1}i_{2}...i_{n}}}}\bra{\psi_{_{i_{1}i_{2}...i_{n}}}}=1$,
the determination of ${\bf{r}}_{c}$ reduces to the following
optimization problem\cite{boyd} \be\begin{array}{cc}
\mbox{minimize} &
C_{\gamma}=\sum_{_{i_{1}i_{2}...i_{n}}}q_{_{i_{1}i_{2}...i_{n}}}P_{_{i_{1}i_{2}...i_{n}}}(\gamma)
\\ & 0\leq P_{_{i_{1}i_{2}...i_{n}}}(\gamma)\leq \frac{1}{d_{1}}\\ &
\sum_{_{i_{1}i_{2}...i_{n}}}P_{_{i_{1}i_{2}...i_{n}}}(\gamma)=1.\end{array}\ee
Always the distribution $P_{_{i_{1}i_{2}...i_{n}}}$ satisfies
$0\leq P_{_{i_{1}i_{2}...i_{n}}}(\gamma) \leq \frac{1}{d_{1}}$
for all pure product states (the proof is given in the Appendix
A). One can calculate the distributions
$P_{_{i_{1}i_{2}...i_{n}}}(\gamma)$, consistent with the
aforementioned optimization problem, from the information about
the boundary of feasible region. To achieve the feasible region we
obtain the extreme points corresponding to the product
distributions $P_{_{i_{1}i_{2}...i_{n}}}(\gamma)$ for every given
product states by applying the special conditions on
$q_{_{i_{1}i_{2}...i_{n}}}$'s parameters. $C_{\gamma}$ themselves
are
 functions of the product distributions, and  they are  in turn are
functions of $\gamma$. They are not real variables of $\gamma$
but the product states will be multiplicative. If this feasible
region constructs a polygon by itself, the corresponding boundary
points of the convex hull will minimize exactly  $C_{\gamma}$ in
Eq. (3-18). This problem is called linear programming , and the
simplex method is the easiest way of solving it. If the feasible
region is not a polygon, with the help of tangent planes in this
region at points which are determined either analytically or
numerically one can define new convex hull which is a polygon and
has encircled  the feasible region. The points on the boundary of
the polygon can approximately determine the minimum value
$C_{\gamma}$ from Eq.(3-18), thus the problem is that of a linear
programming again. In general it is difficult to find this region
and solve the optimization problem, thus it is difficult to find
any generic multipartite EW. In the following sections we
consider some simple but important examples which are solved as
linear programming problem.

\section{Bell states diagonal entanglement
witnesses for multi-qubit system}  Here we provide a multi-qubit
entanglement witness. From the previous section one can show that
the Bell states diagonal observable W for multi qubit system is
defined by \be\label{multi2}
W={\bf{r}}\frac{I_{2^{n}}}{2^{n}}+(1-{\bf{r}})\sum_{i_{1},...,i_{n}=0}^{1}q_{i_{1},i_{2},...,i_{n}}\ket{\psi_{i_{1},i_{2},...,i_{n}}}\bra{\psi_{i_{1},i_{2},...,i_{n}}}
,\ee where $\ket{\psi_{i_{1},i_{2},...,i_{n}}}$ is a Bell state:
 \be
\ket{\psi_{i_{1},i_{2},...,i_{n}}}=(\sigma_{z})^{i_{1}}\otimes
(\sigma_{x})^{i_{2}}\otimes...\otimes
(\sigma_{x})^{i_{2}}\ket{\psi_{ _{0}, _{0},..., _{0}}}, \ee with
\be \ket{\psi_{ _{0}, _{0},...,
_{0}}}=\frac{1}{\sqrt{2}}\sum_{i=0}^{1}\ket{i}_{1}\ket{i}_{2}...\ket{i}_{n},
\ee and $\sigma_{z}$ and $\sigma_{x}$ are the Pauli operators.
 This observable is not a positive
operator  and can not be an EW provided that its expectation value
on any product state
$\ket{\gamma}=\ket{\alpha}_{1}\ket{\alpha}_{2}...\ket{\alpha}_{n}$
is positive.

We consider an easy case $q_{ _{00...00}}=0\;,\;q_{_{10...00}}=x$
with all the other $q$'s being equal, i.e.,
$q_{_{i_{1},i_{2},...,i_{n}}}=\frac{1-x}{2(2^{n-1}-1)}$ except for
$i_{1}=i_{2}=...=i_{n}=0$ and $i_{2}=i_{3}=...=i_{n}=0, i_{1}=1$.
Then the observable W reduces to the following form \be
W={\bf{r}}\frac{I_{2^{n}}}{2^{n}}+\frac{(1-{\bf{r}})}{2(2^{n-1}-1)}((1-x)I_{2^{n}}-(1-x)\ket{\psi_{
_{0}, _{0},..., _{0}}}\bra{\psi_{ _{0}, _{0},...,
_{0}}}+((2^n-1)x-1)\ket{\psi_{ _{1}, _{0},..., _{0}}}\bra{\psi_{
_{1}, _{0},..., _{0}}}). \ee We can calculate $C_{\gamma}$  from
the non negativity of $Tr(W\ket{\gamma}\bra{\gamma})$ for a given
product state $\ket{\gamma}$ \be
C_{\gamma}=\frac{1}{2(2^{n-1}-1)}((1-x)-(1-x)P_{_{00...00}}+((2^{n}-1)x-1)P_{_{10...00}}).
\ee According to the definition of product distributions, we have
\be\label{dis}\begin{array}{c}P_{_{00...0}}=\frac{1}{2}\mid
\alpha_{1}\alpha_{2}...\alpha_{n}+\beta_{1}\beta_{2}...\beta_{n}\mid^{2} \\
P_{_{10...0}}=\frac{1}{2}\mid
\alpha_{1}\alpha_{2}...\alpha_{n}-\beta_{1}\beta_{2}...\beta_{n}\mid^{2},\end{array}
\ee where \be\ket{\alpha}_{i}=\left(
\begin{array}{c}\alpha_{i} \\
\beta_{i}\end{array}\right),\;\;\;i=1,2,...,n.\ee  For the most
general given product states $\gamma$, we determine the extreme
allowed values of these product distributions.
Let\be\begin{array}{c}
\ket{\alpha}_{1}=\ket{\alpha}_{2}=...=\ket{\alpha}_{n}=\left(\begin{array}{c}1
\\ 0 \end{array}\right), \end{array}\ee
then the product distributions are \be
 \left\{\begin{array}{c}P_{_{00...00}}=\frac{1}{2}\\
P_{_{10...00}}=\frac{1}{2}\end{array}\right. .\ee Another choice
will make $P_{_{10...0}}$ minimal
\be\ket{\alpha}_{1}=\ket{\alpha}_{2}=...=\ket{\alpha}_{n}=\frac{1}{\sqrt{2}}\left(\begin{array}{c}1
\\ 1 \end{array}\right),\ee
and the product distributions will become
\be \left\{\begin{array}{c}P_{_{10...0}}=\frac{1}{2^{n-1}}\\
P_{_{10...00}}=0\;\;\;\;\;\end{array}\right. . \ee Similarly we
can also make $P_{_{00...0}}$ minimal
\be\begin{array}{cc}\ket{\alpha}_{2}=\ket{\alpha}_{3}=...=\ket{\alpha}_{n}=\frac{1}{\sqrt{2}}\left(\begin{array}{c} 1 \\ 1 \end{array}\right)\;,\;\ket{\alpha}_{1}=\frac{1}{\sqrt{2}}\left(\begin{array}{c} 1 \\ -1 \end{array}\right) & \Rightarrow\left\{\begin{array}{c}P_{_{00...00}}=0\;\;\;\;\\
P_{_{10...00}}=\frac{1}{2^{n-1}}\end{array}\right.\end{array}.\ee
From the definition of the convex function \cite{boyd} we can
show that the convex combination of these distributions provide a
convex region called the  feasible region, where all points in
the interior of this region satisfy the positivity constraint of
$Tr(W\ket{\gamma}\bra{\gamma})$.
 Then we have \be\label{multi1} \left\{\begin{array}{cc}
\mbox{maximize}&
-C_{\gamma}=\frac{1}{2(2^{n-1}-1)}(-(1-x)+(1-x)P_{_{00...00}}-((2^{n}-1)x-1)P_{_{10...00}})\\
\mbox{subject to} &
2P_{_{00...00}}-2P_{_{10...00}}(1-\frac{1}{2^{n-1}})\leq\frac{1}{2^{n-1}}
\\
&
2P_{_{10...00}}-2P_{_{00...00}}(1-\frac{1}{2^{n-1}})\leq\frac{1}{2^{n-1}}
\\ & P_{_{00...00}}\geq 0,P_{_{10...00}}\geq
0.\end{array}\right. \ee Now we must prove that the feasible
region, constructed from the convex of these points, is a polygon.
Let $P_{+}=P_{_{00...0}}$ and $P_{-}=P_{_{10...0}}$. The equation
of the line passing through $(P_{+}=\frac{1}{2^{n-1}},P_{-}=0)$
and $(P_{+}=\frac{1}{2},P_{-}=\frac{1}{2})$ is \be
P_{-}=(\frac{2^{n-1}}{2^{n-1}-2})P_{+}-\frac{1}{2^{n-1}-2}. \ee
Let us further assume   $P_{-}=\lambda P_{+}$. By intersecting
this equation with the one above we get \be\label{dis2}
P_{+}=\frac{1}{2^{n-1}-\lambda(2^{n-1}-2)}.\ee Now if we assume
$\lambda=0$ we arrive at the point
$(P_{+}=\frac{1}{2^{n-1}},P_{-}=0)$, for $\lambda=1$ we conclude
$(P_{+}=\frac{1}{2},P_{-}=\frac{1}{2})$. One can rewrite
Eq.(\ref{dis}) as $$ P_{\pm}=\frac{1}{2}(
\mid\alpha_{1}\mid^2\mid\alpha_{2}\mid^2...\mid\alpha_{n}\mid^2+(1-\mid\alpha_{1}\mid^2)(1-\mid\alpha_{2}\mid^2)...(1-\mid\alpha_{n}\mid^2)$$
\be\pm 2 \mid\alpha_{1}\mid
(1-\mid\alpha_{1}\mid^2)\mid\alpha_{2}\mid(1-\mid\alpha_{2}\mid^2)...\mid\alpha_{n}\mid(1-\mid\alpha_{n}\mid^2)\cos(\phi)).
\ee Thus we write the Lagrangian as \be {\cal
L}=P_{+}+\mu(P_{-}-\lambda P_{+}),\ee where $\mu$ is the Lagrange
multiplier. With $\mid \alpha_{i}\mid=cos{\theta_{i}}$ we maximize
$\cal L$ with respect to $\theta_{i}$'s and $\phi$\be
\left\{\begin{array}{c}\theta_{1}=\theta_{2}=...=\theta_{n}\Rightarrow
\mid \alpha_{1}\mid=\mid \alpha_{2}\mid=...=\mid \alpha_{n}\mid
\\ \phi=0 \end{array}\right.,\ee
such that \be
\tan^{n}{\theta_{i}}=\frac{1-\sqrt{\lambda}}{1+\sqrt{\lambda}},
\ee so that \be
P_{+}=(\frac{2}{1+\sqrt{\lambda}})\frac{1}{(1+(\frac{1-\sqrt{\lambda}}{1+\sqrt{\lambda}})^{\frac{2}{n}})^{n}}.\ee
As we see the  equation for $P_{+}$ is less than the one in
(\ref{dis2}), moreover, this relation indicates the correctness
of the result (4-26)-(4-30). Thus the convex hull is a polygon
and the optimization problem will be converted into the linear
programming one.

 There is no simple
analytical formula for solving  a linear programming, however
there are a variety of very effective methods, including the
simplex method to  solve them. So, minimization solutions of
$C_{\gamma}$ is obtained by  the simplex method\cite{boyd} and we
have (see Appendix C):

{\bf I)}  For {\bf $0 \leq  x \leq \frac{1}{2^{n-1}+1}$} the
extreme points of the feasible region are
$P_{_{00...00}}=P_{_{10...00}}=\frac{1}{2}$ and the minimum value
of $C_{\gamma}$ is defined by $(C_{\gamma})_{min}=\frac{x}{2}$. By
substituting these values in (3-17) we have \be
-\frac{2^{n-1}x}{1-2^{n-1}x}\leq {\bf{r}} \leq 0 \Rightarrow
{\bf{r}}_{c}=-\frac{2^{n-1}x}{1-2^{n-1}x},\ee where ${\bf{r}}_{c}$
is called the critical {\bf{r}}. By substituting ${\bf{r}}_{c}$ in
(\ref{multi2}) this observable has positive expectation value
under any product state, thus  it will be an EW called critical EW
equal to  \be
W_{c}(x)=\frac{1}{2(2^{n-1}-1)}(I_{2^{n}}-\frac{1-x}{1-2^{n-1}x}\ket{\psi_{
_{0}, _{0},..., _{0}}}\bra{\psi_{ _{0}, _{0},...,
_{0}}}+\frac{(2^n-1)x-1}{1-2^{n-1}x}\ket{\psi_{ _{1}, _{0},...,
_{0}}}\bra{\psi_{ _{1}, _{0},..., _{0}}})),\ee which in the
special case where $x=\frac{1}{2^{n}-1}$ the $W_{c}(x)$ reduces to
\be W_{red}=\frac{1}{2(2^{n-1}-1)}(I_{2^{n}}-2\ket{\psi_{ _{0},
_{0},..., _{0}}}\bra{\psi_{ _{0}, _{0},..., _{0}}}), \ee which is
the well known reduction map.

 {\bf II)} For $ \frac{1}{2^{n-1}+1}\leq x \leq 1$ the extreme
points of the feasible region are
$P_{_{00...00}}=\frac{1}{2^{n-1}}$ and $P_{_{10...00}}=0$
respectively. Therefore,  from the simplex method we get
$(C_{\gamma})_{min}=\frac{1-x}{2^{n}}$, hence
${\bf{r}}_{c}=-\frac{1-x}{x}$ and the critical EW is calculated to
be \be
W_{c}(x)=\frac{1}{(2^{n-1}-1)}(\frac{1-x}{x}\frac{I_{2^{n}}}{2^{n}}-\frac{1-x}{2x}\ket{\psi_{
_{0}, _{0},..., _{0}}}\bra{\psi_{ _{0}, _{0},...,
_{0}}}+\frac{(2^n-1)x-1}{2x}\ket{\psi_{ _{1}, _{0},...,
_{0}}}\bra{\psi_{ _{1}, _{0},..., _{0}}})).\ee Note that this
choice of q is not the only way of defining a BDEW for multi-qubit
system in the one parameter representation. Let us consider the
alternative definition for the  one parameter BDEW by studying the
following example. Assume  $q_{_{00...01}}=x$ and set all  the
other $q$'s to be equal. Thus  we have \be
W={\bf{r}}\frac{I_{2^{n}}}{2^{n}}+\frac{(1-{\bf{r}})}{2(2^{n-1}-1)}((1-x)I_{2^{n}}-(1-x)\ket{\psi_{
_{0}, _{0},..., _{0}}}\bra{\psi_{ _{0}, _{0},...,
_{0}}}+((2^n-1)x-1)\ket{\psi_{ _{0}, _{0},..., _{1}}}\bra{\psi_{
_{0}, _{0},..., _{1}}}). \ee Similarly  we can find the extreme
points of $P_{_{00...00}}$ and $P_{_{00...01}}$ as
\be\begin{array}{cc}
\ket{\alpha}_{1}=\ket{\alpha}_{2}=...=\ket{\alpha}_{n}=\left(\begin{array}{c}1 \\ 0 \end{array}\right) & \left\{\begin{array}{c}P_{_{00...00}}=\frac{1}{2}\\
P_{_{00...01}}=0\end{array}\right.\end{array}.\ee
\be\begin{array}{cc}\ket{\alpha}_{1}=\ket{\alpha}_{2}=...=\ket{\alpha}_{n-1}=\left(\begin{array}{c} 1 \\ 0 \end{array}\right)\;\mbox{and}\;\ket{\alpha}_{n}=\left(\begin{array}{c} 0 \\ 1 \end{array}\right) & \left\{\begin{array}{c}P_{_{00...00}}=0\\
P_{_{00...01}}=\frac{1}{2}\end{array}\right.\end{array}.\ee

 Also
we know that the convex combination of $P_{_{00...00}}$ and
$P_{_{00...01}}$ provides a convex region or a feasible region.
Then we have an optimization problem as follows: \be
\left\{\begin{array}{cc} \mbox{minimize}&
C_{\gamma}=\frac{1}{2(2^{n-1}-1)}((1-x)-(1-x)P_{_{00...00}}+((2^{n}-1)x-1)P_{_{00...01}})\\
\mbox{subject to} & \frac{1}{2}-P_{_{00...00}}-P_{_{00...01}}\geq
0 \\ & P_{_{00...00}},P_{_{00...01}}\geq 0\end{array}\right..\ee
Here the optimization is converted to the linear programming
problem. To prove, we must show that the feasible region is a
polygon. Let us suppose $P_{+}=P_{_{00...00}}$ and
$P_{-}=P_{_{00...01}}$ with \be\left\{\begin{array}{c}
P_{+}=\frac{1}{2}\mid
\alpha_{1}\alpha_{2}...\alpha_{n}+\beta_{1}\beta_{2}...\beta_{n}\mid^{2}
\\ P_{-}=\frac{1}{2}\mid
\beta_{1}\alpha_{2}...\alpha_{n}+\alpha_{1}\beta_{2}...\beta_{n}\mid^{2},\end{array}\right.\ee
where as before we have $P_{-}=\lambda P_{+}$ and $\mid
\alpha_{i}\mid=\cos{\theta_{i}}$. By maximizing the Lagrangian we
get $\theta_{2}=\theta_{3}=...=\theta_{n}$. Thus we
have\be\label{des3}
P_{+}+P_{-}=\frac{1}{2}(\cos^{2n-2}{\theta_{2}}+\sin^{2n-2}{\theta_{2}})\leq
\frac{1}{2}. \ee The line passing through the points
$(P_{+}=\frac{1}{2},P_{-}=0)$ and $(P_{+}=0,P_{-}=\frac{1}{2})$
is $P_{-}=-P_{+}+\frac{1}{2}$, which is always located above the
curve obtained in the Eq.(\ref{des3}).  Therefore, all the points
are within the feasible region and this region constructs a
polygon. Thus the above optimization problem reduces to a linear
programming one. This minimization is exactly solved in the same
way as mentioned above, and the critical EW is obtained as \be
{\bf{r}}_{c}=\frac{-2^{n}(1-x)}{2(2^{n-1}(x+1)-2)}\Rightarrow
\ee$$
W_{c}=\frac{1}{(2^{n-1}(x+1)-2)}((1-x)I_{2^{n}}-2(1-x)\ket{\psi_{_{00..00}}}\bra{\psi_{_{00..00}}}+2((2^{n}-1)x-1)\ket{\psi_{_{00..01}}}\bra{\psi_{_{00..01}}}).$$

\section{Bell states diagonal entanglement
witnesses for $2 \otimes N$ system}  Here, we will find a $2
\otimes N$ entanglement witness. From the previous discussions we
can define the Bell states diagonal observable  W as \be
W={\bf{r}}\frac{I_{2N}}{2N}+(1-{\bf{r}})\sum_{i=0}^{N-1}\sum_{\alpha=0}^{1}q_{_{i\alpha}}\ket{\psi_{_{i\alpha}}}\bra{\psi_{_{i\alpha}}}
,\ee where $\ket{\psi_{_{i\alpha}}}=I_{2}\otimes
(S)^{i}(\Omega)^{\alpha}\ket{\psi_{_{00}}}$, with
$\ket{\psi_{_{00}}}=\frac{1}{\sqrt{2}}\sum_{k=0}^{1}\ket{k}\ket{k}$
and \be
\begin{array}{cc} \omega=\left(\begin{array}{ccccc} 1 & 0 & 0 & \cdots &
0\\0 & -1 & 0 & \cdots & 0\\0 & 0 & 1 & \cdots & 0 \\
\vdots & &  \ddots & & \vdots\\0 & 0 & 0 & \cdots & 1
\end{array}\right) & S=\left(\begin{array}{ccccc} 0 & 1 & 0 & \cdots &
0\\1 & 0 & 0 & \cdots & 0\\0 & 0 & 1 & \cdots & 0 \\
\vdots & &  \ddots & & \vdots\\0 & 0 & 0 & \cdots & 1
\end{array}\right).\end{array} \ee
Similar to multi-qubit let $q_{_{00}}=0$ and $q_{_{10}}=x$ and
let all the other q's be equal to $\frac{1-x}{2N-2}$. Then by
obtaining the expectation value of W on the product states and
finding the product distributions we have \be\begin{array}{cc}
\ket{\alpha}_{1}=\left(\begin{array}{c}1 \\ 0 \end{array}\right)\;\;,\;\;\ket{\alpha}_{2}=\left(\begin{array}{c}1 \\ 0\\ \vdots \\0 \end{array}\right) & \left\{\begin{array}{c}P_{_{00}}=\frac{1}{2}\\
P_{_{10}}=\frac{1}{2}\end{array}\right.\end{array}\ee
\be\begin{array}{cc}\ket{\alpha}_{1}=\frac{1}{\sqrt{2}}\left(\begin{array}{c}1
\\ 1 \end{array}\right)\;\;,\;\;\ket{\alpha}_{2}=\frac{1}{\sqrt{2}}\left(\begin{array}{c}1 \\ 1\\ 0\\ \vdots \\0 \end{array}\right)
& \left\{\begin{array}{c}P_{_{00}}=\frac{1}{2}\\
P_{_{10}}=0\end{array}\right.\end{array}\ee
\be\begin{array}{cc}\ket{\alpha}_{1}=\frac{1}{\sqrt{2}}\left(\begin{array}{c}
1 \\ -1
\end{array}\right)\;\;,\;\;\ket{\alpha}_{2}=\frac{1}{\sqrt{2}}\left(\begin{array}{c}1
\\ 1\\ 0\\ \vdots \\0 \end{array}\right)
 & \left\{\begin{array}{c}P_{_{00}}=0\\
P_{_{10}}=\frac{1}{2}\end{array}\right.\end{array}.\ee The
feasible region is a rectangular and the optimization problem
reduces to linear programming. Therefore, by using simplex method
for $ 0 \leq x\leq\frac{1}{N+1}$ we find the  minimum value of
$C_{\gamma}=\frac{x}{2}$ and critical ${\bf{r}}$ as
${\bf{r}}_{c}=\frac{-Nx}{1-Nx}$, and the critical EW is defined as
\be
W_{c}=\frac{1}{2(N-1)}(I_{2N}-\frac{1-x}{1-Nx}\ket{\psi_{_{00}}}\bra{\psi_{_{00}}}+\frac{(2N-1)x-1}{1-Nx}\ket{\psi_{_{10}}}\bra{\psi_{_{10}}}).
\ee For the  critical ${\bf{r}}$ we find
${\bf{r}}_{c}=-\frac{1-x}{x}$ in the region  $ \frac{1}{N+1} \leq
x\leq 1$ and the critical EW has the following form \be
W_{c}(x)=\frac{1}{2}(\frac{1-x}{x}\frac{I_{2N}}{2N}-\frac{1-x}{2x}\ket{\psi_{_{00}}}\bra{\psi_{_{00}}}+\frac{(2N-1)x-1}{2x}\ket{\psi_{_{10}}}\bra{\psi_{_{10}}}).\ee

In another one parameter EW  example we assume that $q_{_{01}}=x$
and set all the other $q$'s to be equal so that we have \be
W={\bf{r}}\frac{I_{2N}}{2N}+\frac{(1-{\bf{r}})}{2(N-1)}((1-x)I_{2N}-(1-x)\ket{\psi_{_{00}}}\bra{\psi_{_{00}}}+((2N-1)x-1)\ket{\psi_{
_{01}}}\bra{\psi_{_{01}}}). \ee Similarly  we can find the
extreme points of $P_{_{00}}$ and $P_{_{01}}$ as
\be\begin{array}{cc}\ket{\alpha}_{1}=\left(\begin{array}{c}1
\\ 0 \end{array}\right)\;\;,\;\;\ket{\alpha}_{2}=\left(\begin{array}{c}1 \\ 0\\ \vdots \\0 \end{array}\right)
& \left\{\begin{array}{c}P_{_{00}}=\frac{1}{2}\\
P_{_{01}}=0\end{array}\right.\end{array},\ee
\be\begin{array}{cc}\ket{\alpha}_{1}=\left(\begin{array}{c} 1
\\ 0
\end{array}\right)\;\;,\;\;\ket{\alpha}_{2}=\left(\begin{array}{c}0
\\ 1\\ 0\\ \vdots \\0 \end{array}\right)
 & \left\{\begin{array}{c}P_{_{00}}=0\\
P_{_{01}}=\frac{1}{2}\end{array}\right.\end{array}.\ee Then the
critical EW is defined as \be
{\bf{r}}_{c}=\frac{-2N(1-x)}{2(N(x+1)-2)}\Rightarrow \ee
$$W_{c}=\frac{1}{(N(x+1)-2)}((1-x)I_{2N}-(1-x)\ket{\psi_{_{00}}}\bra{\psi_{_{00}}}+((2N-1)x-1)\ket{\psi_{_{01}}}\bra{\psi_{_{01}}}).$$
\section{Bell states diagonal entanglement
witnesses for $3 \otimes 3$ system} Here we provide a  $3 \otimes
3$ Bell diagonal entanglement witness. One can show that the Eq.
(\ref{GWE1}) for a $3 \otimes 3$ system reads as \be
W={\bf{r}}\frac{I_{9}}{9}+(1-{\bf{r}})\sum_{i_{1},i_{2}=0}^{2}q_{i_{1},i_{2}}\ket{\psi_{i_{1},i_{2}}}\bra{\psi_{i_{1},i_{2}}}.\ee
It is difficult to prove  whether or not the EW for a $3 \otimes
3$ system is optimal. Also it is difficult to see for which value
of the allowed ${\bf{r}}$, EW are (or are not) decomposable.
Therefore, to investigate the optimality  and non decomposability
of these EW we
restrict ourselves below  to some particular choice of $q_{ij}$:\\
Because the distributions $0\leq P_{ij}\leq\frac{1}{3}$ and the
minimum value of  $C_{\gamma}$
 are dependent on the  coefficients $q_{ij}$, we consider a special
 case for
the  coefficients $q_{ij}$ defined  by \be
q_{01}=q_{02}=q_{11}=q_{22}=q_{12}=q_{21}=\frac{1}{8}\;\;,\;\;
q_{10}=x\;\; \mbox{and}\;\; q_{20}=\frac{1}{4}-x\;\;,\;\;0\leq
x\leq\frac{1}{4}. \ee By substituting these values in (\ref{GWE1})
we get \be\label{ciri1}
W(x)={\bf{r}}\frac{I_{9}}{9}+(1-{\bf{r}})(\frac{I_{9}}{8}-\frac{1}{8}\ket{\psi_{_{00}}}\bra{\psi_{_{00}}}-\frac{8x-1}{8}(\ket{\psi_{_{10}}}\bra{\psi_{_{10}}}-\ket{\psi_{_{20}}}\bra{\psi_{_{20}}}))
.\ee By using (\ref{optt1}) for non-negativity of the observable W
we find the distributions $P_{ij}$ as a function of x. The minimum
value of $C_{\gamma}$  is obtained  from the boundary of the
feasible region, i.e., we have \be
C_{\gamma}=\frac{1}{8}(1-P_{_{00}}-(8x-1)(P_{_{10}}-P_{_{20}})).
\ee  For given product states
$\ket{\gamma}=\ket{\alpha}_{1}\ket{\alpha}_{2}$ one can obtain
the extreme points of the product distributions as
$$\left\{\begin{array}{ccc}
\ket{\alpha}_{1}=\ket{\alpha}_{2}=\frac{1}{\sqrt{3}}\left(\begin{array}{c}1\\1\\1\end{array}\right)
& \rightarrow & (P_{00}=\frac{1}{3},P_{10}=0,P_{20}=0)
\\ \ket{\alpha}_{1}=\ket{\alpha}_{2}=\frac{1}{\sqrt{3}}\left(\begin{array}{c}1\\ \omega\\ \bar{\omega}\end{array}\right)
& \rightarrow & (P_{00}=0,P_{10}=\frac{1}{3},P_{20}=0) \\
\ket{\alpha}_{1}=\ket{\alpha}_{2}=\frac{1}{\sqrt{3}}\left(\begin{array}{c}1\\
\bar{\omega}\\ \omega\end{array}\right) & \rightarrow &
(P_{00}=0,P_{10}=0,P_{20}=\frac{1}{3})\end{array}\right.$$
\be\left\{\begin{array}{ccc}
\ket{\alpha}_{1}=\frac{1}{\sqrt{2}}\left(\begin{array}{c}0\\1\\1\end{array}\right),
\ket{\alpha}_{2}=\frac{1}{\sqrt{2}}\left(\begin{array}{c}0\\-1\\1\end{array}\right)
& \rightarrow & (P_{00}=0,P_{10}=\frac{1}{4},P_{20}=\frac{1}{4})
\\ \ket{\alpha}_{1}=\frac{1}{\sqrt{2}}\left(\begin{array}{c}1\\ 0\\ \bar{\omega}\end{array}\right), \ket{\alpha}_{2}=\frac{1}{\sqrt{2}}\left(\begin{array}{c}1\\ 0\\ -\bar{\omega}\end{array}\right)
& \rightarrow & (P_{00}=\frac{1}{4},P_{10}=0,P_{20}=\frac{1}{4}) \\
\ket{\alpha}_{1}=\frac{1}{\sqrt{2}}\left(\begin{array}{c}1\\
\omega\\ 0\end{array}\right), \ket{\alpha}_{2}=\frac{1}{\sqrt{2}}\left(\begin{array}{c}1\\
-{\omega}\\ 0\end{array}\right) & \rightarrow &
(P_{00}=\frac{1}{4},P_{10}=\frac{1}{4},P_{20}=0)
\end{array}\right.\ee and
\be\begin{array}{ccc} \ket{\alpha}_{1}=\ket{\alpha}_{2}=\left(\begin{array}{c}1\\
0 \\ 0 \end{array}\right) & \rightarrow &
(P_{00}=\frac{1}{3},P_{10}=\frac{1}{3},P_{20}=\frac{1}{3})\end{array}
.\ee By convex combination of these points we obtain the possible
region. Thus we have an optimization problem as  \be\label{HJ}
\left\{\begin{array}{cc} \mbox{minimize}&
C_{\gamma}=\frac{1}{8}(1-P_{_{00}}-(8x-1)(P_{_{10}}-P_{_{20}}))\\
\mbox{subject to} & 1-3P_{_{00}}-P_{_{10}}+P_{_{20}}\geq 0
\\ &
1+P_{_{00}}-3P_{_{10}}-P_{_{20}}\geq 0 \\ &
1-P_{_{00}}+P_{_{10}}-3P_{_{20}}\geq 0 \\ &
P_{_{00}},P_{_{10}},P_{_{20}}\geq 0.\end{array}\right. \ee One
can prove analytically that the region can be encircled  with a
polygon and the optimization problem is reduced to a linear
programming. To prove, we begin from the definition of the product
distributions \be
\begin{array}{c}P_{00}=\frac{1}{3}\mid
\alpha_{1}\beta_{1}+\alpha_{2}\beta_{2}+\alpha_{3}\beta_{3}\mid
^2 \\ P_{10}=\frac{1}{3}\mid
\alpha_{1}\beta_{1}+\alpha_{2}\beta_{2}\omega+\alpha_{3}\beta_{3}\bar{\omega}\mid
^2\\ P_{20}=\frac{1}{3}\mid
\alpha_{1}\beta_{1}+\alpha_{2}\beta_{2}\bar{\omega}+\alpha_{3}\beta_{3}{\omega}\mid
^2.\end{array}\ee Without loss of generality, one can assume that
\be \mid \alpha_{1}\beta_{1}\mid=x_{1}\;\;,\;\;\mid
\alpha_{2}\beta_{2}\mid=x_{2}\;\;,\;\;\mid
\alpha_{3}\beta_{3}\mid=x_{3}.\ee The Schwartz inequality yields
\be\label{PI} x_{1}+x_{2}+x_{3}\leq 1. \ee Since we are looking
the extreme points we will choose the maximum value in the
inequality (\ref{PI}), that is $ x_{1}+x_{2}+x_{3}=1.$ Thus the
product distributions are written as \be
\begin{array}{c}P_{00}=\frac{1}{3}\mid
x_{1}+x_{2}e^{i\phi_{2}}+x_{3}e^{i\phi_{3}}\mid ^2 \\
P_{10}=\frac{1}{3}\mid
x_{1}+x_{2}e^{i\phi_{2}}\omega+x_{3}e^{i\phi_{3}}\bar{\omega}\mid ^2\\
P_{20}=\frac{1}{3}\mid
x_{1}+x_{2}e^{i\phi_{2}}\bar{\omega}+x_{3}e^{i\phi_{3}}\omega\mid
^2.\end{array}\ee Now, supposing that $P_{00}$ and $P_{10}$ are
fixed values, we conclude \be \phi_{2}=\phi_{3}\;\;,\;\;
x_{2}=x_{3}. \ee Thus
\be\begin{array}{c}P_{00}=\frac{1}{3}\mid 1+2x_{2}(1-\cos{\phi_{2}})\mid ^2 \\
P_{10}=\frac{1}{3}\mid 1+2x_{2}(1-\cos{\phi_{2}+\frac{2\pi}{3}})\mid ^2\\
P_{20}=\frac{1}{3}\mid
1+2x_{2}(1-\cos{\phi_{2}-\frac{2\pi}{3}})\mid ^2.\end{array}\ee
We write down the equation for the planes passing through the
obtained extreme points for $P_{ij}$ and maximize it with respect
to the variables $\phi_{2}$ and $x_{2}$. Where the obtained
values for $\phi_{2}$ and $x_{2}$ are indicative of the violation
from the equations of planes, that is there are points which are
located out of these planes, or in other word the planes have
become convex.  Thus, the equations of the planes (Fig 1) and
their maximum violation(D) are obtained as follows
\be\left\{\begin{array}{ccc}1)3P_{10}-P_{20}+P_{00}-1=0 &
x_{2}=\frac{7}{61}, \cos{\phi_{2}}=\frac{-1}{7} & D=\frac{2}{61}\\
2)3P_{10}+P_{20}-P_{00}-1=0 & x_{2}=\frac{7}{61},
\cos{\phi_{2}}=\frac{-11}{14} &
D=\frac{2}{61}\\3)3P_{20}+P_{10}-P_{00}-1=0 & x_{2}=\frac{7}{61},
\cos{\phi_{2}}=\frac{-11}{14} &
D=\frac{2}{61}\\4)3P_{20}-P_{10}+P_{00}-1=0 & x_{2}=\frac{7}{61},
\cos{\phi_{2}}=\frac{-1}{7} &
D=\frac{2}{61}\\5)3P_{00}+P_{20}-P_{10}-1=0 & x_{2}=\frac{7}{61},
\cos{\phi_{2}}=\frac{13}{14} &
D=\frac{2}{61}\\6)3P_{00}-P_{20}+P_{10}-1=0 & x_{2}=\frac{7}{61},
\cos{\phi_{2}}=\frac{13}{14} & D=\frac{2}{61}.\end{array}\right.
\ee It is seen that the points thus obtained are located out of
the considered plane. Thus, the equations of the planes passing
through the new extreme points which are parallel to the above
plane are obtained. For example, the  equation of plane parallel
to $3p_{00}+P_{10}+p20=1$ is $3p_{00}+P_{10}+p20=1+\frac{2}{61}$
which under permutation $(P_{00},P_{10},P_{20})$ will act
similarly. Tack  arbitrary any three of planes passing through
the new extreme,
$(P_{00}=\frac{1}{3},P_{10}=\frac{1}{3},P_{20}=\frac{1}{3})$,
$(P_{10}=0, P_{20}=0, P_{00}=0)$ and $P_{10}+P_{20}+P_{00}-1=0$
and intersecting with each other. Hence new extreme points will
be produced. Thus we have encircled  a polygon by its feasible
region and the optimization problem will be reduced to that of a
linear programming.  The vertices of this polygon are the
solutions of the problem provided that they obey
$(P_{00}\leq\frac{1}{3},P_{10}\leq\frac{1}{3},P_{20}\leq\frac{1}{3})$
and by substituting them into equation $C_{\gamma}$ one can
determine its minimum value. Thus, for $\frac{67}{756}\leq x\leq
\frac{61}{378}$, the extreme point is defined by
$(\frac{1}{3},\frac{1}{3},\frac{1}{3})$ and finally
 $(C_{\gamma})_{min}=(\frac{1}{12})$.  Having found  the critical ${\bf{r}}$ we  substitute it in (\ref{GWE1}) and obtain a family of EW (called
critical EW). Thus we have \be\label{op8}
p_{c}=-3\;,\;W_{c}(x)=\frac{1}{2}(\frac{1}{3}I_{9}-\ket{\psi_{_{00}}}\bra{\psi_{_{00}}}+(8x-1)(\ket{\psi_{_{10}}}\bra{\psi_{_{10}}}-\ket{\psi_{_{20}}}\bra{\psi_{_{20}}})),\ee
where  $W_{c}(x)$ reduces to the following well known reduced EW
at $x=\frac{1}{8}$: \be\label{red}
W_{red}=\frac{I_{9}-3\ket{\psi_{00}}\bra{\psi_{00}}}{6}, \ee In
the Appendix B  it is shown that the above EW is optimal in
contrast to the conclusion that it is a decomposable EW
Ref.\cite{horodecki4}.

In the Appendix B, we discuss the possible choice of  x consistent
with $C_{mn}=\frac{1}{12}$ and the optimality of the
corresponding $W_{c}(x)$.

Also  we  prove in the following section that $W_{c}$ is nd-EW
for all values of $\frac{67}{756}\leq x\leq \frac{61}{378}$,
except for $x=\frac{1}{8}$. Besides taking a convex combination of
$W_{c}$ and $W_{red}$ ,i.e., \be\label{prime} W_{\Lambda}=\Lambda
W_{c}+(1-\Lambda)W_{red},\ee we obtain a new EW which is optimal
(see Appendix B) and is also an nd-EW for certain value of the
parameter $\Lambda$  as  will be shown in section 7.

However  we can consider other values for $q_{ij}$ in
(\ref{GWE1}), e.g.,
$q_{20}=q_{02}=q_{11}=q_{22}=q_{12}=q_{21}=\frac{1}{8}\;,\;
q_{10}=x$ \mbox{and} $q_{01}=\frac{1}{4}-x\;,\;0\leq
x\leq\frac{1}{4}$ then  define the observable W by substituting
the above condition in (\ref{GWE1}) as follows \be\label{ciri1}
W(x)={\bf{r}}\frac{I_{9}}{9}+(1-{\bf{r}})(\frac{I_{9}}{8}-\frac{1}{8}\ket{\psi_{_{00}}}\bra{\psi_{_{00}}}-\frac{8x-1}{8}(\ket{\psi_{_{10}}}\bra{\psi_{_{10}}}-\ket{\psi_{_{01}}}\bra{\psi_{_{01}}}))
.\ee
 By using (\ref{optt1}) for non-negativity of the observable W
we find the distributions $P_{ij}$ as a function of x. The minimum
value of $C_{\gamma}$  is obtained  from the boundary of the
feasible region, i.e., we have \be
C_{\gamma}=\frac{1}{8}(1-P_{_{00}}-(8x-1)(P_{_{10}}-P_{_{01}})).\ee
 For given product states
$\ket{\gamma}=\ket{\alpha}_{1}\ket{\alpha}_{2}$ one can obtain
the extreme points of the product distributions as
$$\left\{\begin{array}{ccc}
\ket{\alpha}_{1}=\ket{\alpha}_{2}=\left(\begin{array}{c}1\\0\\0\end{array}\right)
& \rightarrow & (P_{00}=\frac{1}{3},P_{10}=\frac{1}{3},P_{01}=0)
\\ \ket{\alpha}_{1}=\ket{\alpha}_{2}=\frac{1}{\sqrt{3}}\left(\begin{array}{c}1\\ 1\\ 1\end{array}\right)
& \rightarrow & (P_{00}=\frac{1}{3},P_{10}=0,P_{01}=\frac{1}{3}) \\
\ket{\alpha}_{1}=\frac{1}{\sqrt{3}}\left(\begin{array}{c}\omega\\
\omega\\ \bar{\omega}\end{array}\right)\;,\;\ket{\alpha}_{2}=\frac{1}{\sqrt{3}}\left(\begin{array}{c}\bar{\omega}\\
{\omega}\\ \bar{\omega}\end{array}\right) & \rightarrow &
(P_{00}=0,P_{10}=\frac{1}{3},P_{01}=\frac{1}{3})\end{array}\right.$$
\be\left\{\begin{array}{ccc}
\ket{\alpha}_{1}=\frac{1}{\sqrt{3}}\left(\begin{array}{c}1\\1\\
\omega\end{array}\right),
\ket{\alpha}_{2}=\frac{1}{\sqrt{3}}\left(\begin{array}{c}1\\1\\
\bar{\omega}\end{array}\right) & \rightarrow &
(P_{00}=\frac{1}{3},P_{10}=0,P_{01}=0)
\\ \ket{\alpha}_{1}=\frac{1}{\sqrt{3}}\left(\begin{array}{c}1\\ \omega\\ \bar{\omega}\end{array}\right), \ket{\alpha}_{2}=\frac{1}{\sqrt{3}}\left(\begin{array}{c}1\\ \omega\\ \bar{\omega}\end{array}\right)
& \rightarrow & (P_{00}=0,P_{10}=\frac{1}{3},P_{01}=0) \\
\ket{\alpha}_{1}=\frac{1}{\sqrt{3}}\left(\begin{array}{c}1\\
\omega\\ 1\end{array}\right), \ket{\alpha}_{2}=\frac{1}{\sqrt{3}}\left(\begin{array}{c}1\\
1\\ \bar{\omega}\end{array}\right) & \rightarrow &
(P_{00}=0,P_{10}=0,P_{01}=\frac{1}{3})\end{array}\right. .\ee
Similar to  $(P_{00},P_{10},P_{20})$ we obtain the extreme
points. Then  by convex combination of these points we obtain the
feasible region (see Fig 2). The optimization problem is in the
following form \be\label{hj} \left\{\begin{array}{cc}
\mbox{minimize}&
C_{\gamma}=\frac{1}{8}(1-P_{_{00}}-(8x-1)(P_{_{10}}-P_{_{01}})\\
\mbox{subject to} & \frac{2}{3}-P_{00}-P_{10}-P_{01}\geq 0
\\ & P_{_{00}},P_{_{10}},P_{_{01}}\geq 0.\end{array}\right. \ee
We have been able to fined analytically the extreme points which
at the same time don't violate the plane
$(\frac{2}{3}-P_{00}-P_{10}-P_{01}=0)$. But we have failed to
prove in general that no point lies out of the plane. Therefore,
we have proved numerically that there is no violation from the
plane. Thus, this feasible region is a convex hull or a polygon
itself and reduces the optimization problem to a linear
programming. So the vertices of the polygon are the solutions of
the problem which by substituting them into equation $C_{\gamma}$
one can determine its minimum value as
$(C_{\gamma})_{min}=(\frac{2-(8x-1)}{24})$. We can find the
critical r and by substituting the critical r in (\ref{GWE1}) we
obtain a family of EW (called critical EW) resulting in
$${\bf{r}}_{c}=\frac{-3+24x}{-1+24x}$$
\be
W_{c}(x)=\frac{1}{3(-1+24x))}((8x-1))I_{9}-3\ket{\psi_{_{00}}}\bra{\psi_{_{00}}}-3(8x-1)(\ket{\psi_{_{10}}}\bra{\psi_{_{10}}}-\ket{\psi_{_{01}}}\bra{\psi_{_{01}}})).\ee

The obtained EW for two sets of triplet P, namely
$(P_{00},P_{20},P_{20})$ and $(P_{00},P_{20},P_{01})$ can produce
the most general form EW corresponding to combination of another
triplet $P_{ij}$'s. Since under Fourier  transform one can
transform all the shifts in to the modulation. Moreover, the
shift and modulation operators themselves can affect on the
corresponding EW too, and so produce new combination of triplet.
\section{Non-decomposible 3 $\otimes$ 3 Bell states diagonal entanglement witnesses}
By calculating the partial transpose of $W_{c}(\frac{67}{756}\leq
x\leq \frac{61}{378})$ (for $\{P_{00},P_{10},P_{20}$ case) we
prove that  it is an nd-EW. The necessary and sufficient
condition for non-decomposibility of $W_{c}$ reduces to the
negativity of its partial transpose. Using the following relation
\be\label{pt} (\mid
\psi_{j^{\prime}k^{\prime}}><\psi_{jk}\mid)^{T_{A}}=\frac{1}{3}\sum_{l,m}\omega^{ml}\mid\psi_{m+j^{\prime},l+k^{\prime}}><\psi_{m+j,3-(l-k)}\mid,\ee
one can show that $(W_{c})^{T_{A}}$ is a block diagonal, i.e., we
have $$(W_{c})^{T_{A}}=\sum_{j,k,k^{\prime}}
(O_{j})_{kk^{\prime}}\mid
\psi_{j^{\prime}k^{\prime}}><\psi_{jk}\mid,$$ with the matrices
$O_{j}$ calculated as
\be O_{j}=\left( \begin{array}{ccc} 0 & 0 & 0 \\
0 & \frac{1}{6} &
C_{j} \\
0 & \bar{C_{j}} & \frac{1}{6}
\end{array}\right),\ee
with \be\label{AA}
C_{j}=\frac{4}{3}(x\omega+(\frac{1}{4}-x)\bar{\omega})\bar{\omega}^{j}\;\;,\;\;j=0,1,2.\ee
Using the fact that $\mid C_{2}\mid=\mid C_{1}\mid=\mid C_{0}\mid
$ one can show that the matrices $O_{j}$ have the same eigenvalues
\be\label{nd2}\left\{ \begin{array}{c} \lambda=0 \\
\lambda_{\pm}=\lambda^{j}_{\pm}=\frac{1}{6}\pm
\frac{1}{6}\sqrt{4+48x(4x-1)}.\end{array}\right. \ee The above
equation indicates that $\lambda_{-}$ is negative except for the
particular case in which  $x=\frac{1}{8}$, i.e., $W_{red}$. Then
different eigenvectors are so obtained \be\left\{
\begin{array}{ccc} \lambda=0 & \rightarrow
\mid\phi^{0}_{j}>=\mid\psi_{j0}>,
\\
\lambda=\lambda_{\pm} & \rightarrow
\mid\phi^{\pm}_{j}>=\frac{1}{\sqrt{\mid\beta_{\pm}^{j}\mid^{2}+1}}(\beta^{j}_{\pm}\mid\psi_{j1}>+\mid\psi_{j2}>),
\end{array}\right. ,\ee
where \be
\beta^{j}_{\pm}=\frac{C_{j}\lambda_{\pm}}{\lambda_{\pm}^{2}-\lambda_{\pm}B}.
\ee  So we conclude that $W_{c}^{T_{A}}$ has three eigenvalues,
namely $\lambda_{0},\lambda_{\pm}$,  each with degeneracy 3, and
the following projection operators
 \be\label{pro} \left\{
\begin{array}{c}
Q_{+}=\sum_{j=0}^{2}\mid\phi_{j}^{+}><\phi_{j}^{+}\mid \\
Q_{-}=\sum_{j=0}^{2}\mid\phi_{j}^{-}><\phi_{j}^{-}\mid \\
Q_{0}=\sum_{j=0}^{2}\mid\phi^{0}_{j}><\phi^{0}_{j}\mid.
\end{array}\right. \ee Here we have \be\label{optimalew}
W_{c}^{T_{A}}=\lambda_{+}Q_{+}-\mid\lambda_{-}\mid Q_{-}. \ee The
 equation indicates that $W_{c}^{T_{A}}$ is not a positive
definite operator except for the particular case $W_{red}$, hence
it is non-decomposable entanglement witness.

We are interested  in the n-d of EW given in (\ref{GWE1}) for the
allowed values of p. Therefore, we write Eq.(\ref{GWE1}) as \be
\label{nd3} W=\varepsilon I_{9}/9 +(1-\varepsilon)W_{c}, \ee
with\be\label{epsilon} \varepsilon=\frac{{\bf{r}}+3}{4}. \ee Now,
expanding $I_{9}/9$ in terms of the projection operator
(\ref{pro}) as \be
I_{9}=Q^{T_{A}}_{0}+Q^{T_{A}}_{-}+Q^{T_{A}}_{+}, \ee the EW given
by (\ref{GWE1}) can be written as\be\label{nd6} W=\varepsilon/9
Q^{T_{A}}_{0}+(\frac{\varepsilon}{9}+(1-\varepsilon)\lambda_{+})Q^{T_{A}}_{+}+(\frac{\varepsilon}{9}-(1-\varepsilon)\mid\lambda_{-}\mid)Q^{T_{A}}_{-}.
\ee The above form of EW indicates that its partial transpose
$W^{T_{A}}$ is positive, i.e., it is  decomposable EW if we have
\be\label{nd4} W^{T_{A}}\geq
0\Rightarrow(\frac{\varepsilon}{9}-(1-\varepsilon)\mid\lambda_{-}\mid)\geq
0 \rightarrow {\bf{r}}\geq \frac{-3+9\mid \lambda_{-}\mid}{1+9\mid
\lambda_{-}\mid},\ee for $\frac{-3+9\mid \lambda_{-}\mid}{1+9\mid
\lambda_{-}\mid}\leq {\bf{r}} \leq -3$. It is not easy  to tell
where the EW is or is not  decomposable. In the next section
using some bound entangled state we will investigate their
non-decomposability.

Now, in the remaining part of this section we try to obtain some
nd-EW by taking the convex combination $W_{c}(x)\;\; \mbox{for all
}\;\;\frac{67}{756}\leq x\leq \frac{61}{378}$  and $W_{red}$
(\ref{red}) as \be\label{prime} W_{\Lambda}(x)=\Lambda
W_{c}(x)+(1-\Lambda)W_{red}\;\;,\;\;\Lambda\in[0,1].\ee In order
to test the positivity of $W_{\Lambda}^{T_{A}}(x)$ we must first
expand $W_{c}$ and $W_{red}$ in terms of the positive diagonal
operators. Thus at first we write the projection operators
defined in (\ref{pro}) in the following form \be\label{proj}
Q_{\pm}=\sum_{k=0}^{2}\ket{\chi^{\pm}_{k}}\bra{\chi^{\pm}_{k}}\;\;,\;\;Q_{0}=\sum_{k=0}^{2}\ket{\psi_{ko}}\bra{\psi_{k0}}\ee
with \be\label{vec} \mid\chi^{\pm}_{k}>=(\mid\psi_{k1}>\pm
\omega^{-k}\mid\psi_{k2}>). \ee Now writing $I_{9}/9$ in terms of
the projection operator (\ref{proj}) and using the fact that
$$(\mid\psi_{00}><\psi_{00}\mid)^{T_{A}}=\frac{1}{3}(\sum_{k=0}^{2}\mid\psi_{k0}><\psi_{k0}\mid+\sum_{k=0}^{2}\mid\chi^{+}_{k}><\chi^{+}_{k}\mid-\sum_{k=0}^{2}\mid\chi^{-}_{k}><\chi^{-}_{k}\mid)
$$ and \be
W^{T_{A}}_{c}(x)=\lambda_{+}\sum_{k=0}^{2}\mid\chi^{+}_{k}><\chi^{+}_{k}\mid-\mid\lambda_{-}\mid\sum_{k=0}^{2}\mid\chi^{-}_{k}><\chi^{-}_{k}\mid,
\ee we get for the partial transpose $W_{\Lambda}(x)$in
Eq.(\ref{prime}) \be {W_{\Lambda}}^{T_{A}}(x)=\Lambda
(\lambda_{+})\sum_{k=0}^{2}\mid\chi^{+}_{k}><\chi^{+}_{k}\mid+(-\Lambda
\mid\lambda_{-}\mid+\frac{(1-\Lambda)}{3})\mid\chi^{-}_{k}><\chi^{-}_{k}\mid.
\ee This expression implies that ${W_{\Lambda}}^{T_{A}}(x)$ is
positive, since \be \Lambda \leq
\frac{1}{1+3\mid\lambda_{-}\mid}. \ee  Again, for
$\frac{1}{1+3\mid\lambda_{-}\mid}\leq \Lambda \leq 1 $, it is not
easy to talk about decomposable or non-decomposable
${W_{\Lambda}}(x)$, and one needs to find some bound entangled
states to show their non-decomposability, this will be done in
the following section.
\section{Detection of bound entangled state with Bell states diagonal entanglement witnesses}
Now if we succeed to find any bound entangled
state\cite{woronowicz,horodecki1} so that BDEW is able to detect
this bound state corresponding to BDEW, from definition 2 in
section 1 EW will be an  nd-EW. Let a bound entangled Bell
decomposable state be written  as  \be\label{nd7} \rho=\mu
Q_{0}^{T_{A}}+ \eta Q_{+}^{T_{A}}+\zeta Q_{-}^{T_{A}}\;\;,\;\;
\rho^{T_{A}}\geq 0 \Rightarrow \{\mu,\eta,\zeta\}\geq 0. \ee
Optimal BDEW must detect this bound state, i.e., \be
Tr[W_{c}\rho]<0 \Rightarrow
\eta\lambda_{+}<\zeta\mid\lambda_{-}\mid. \ee On the other hand
this bound state must be positive. For simplicity we use the
operator $W_{c}$ and the identity operator $I_{9}$ in the bound
state definition \be
Q^{T_{A}}_{+}=\frac{W_{c}+\mid\lambda_{-}\mid(I_{9}-Q^{T_{A}}_{0})}{\mid\lambda_{-}\mid
+\lambda_{+}
}\;\;,\;\;Q^{T_{A}}_{-}=\frac{-W_{c}+\lambda_{+}(I_{9}-Q^{T_{A}}_{0})}{\mid\lambda_{-}\mid
+\lambda_{+} },\ee so that the  bound state reduces to the
following form \be\label{nd5}
\rho=(\mu-\frac{\eta\mid\lambda_{-}\mid+\zeta\lambda_{+}}{\mid\lambda_{-}\mid
+\lambda_{+}})Q_{0}^{T_{A}}+(
\frac{\eta\mid\lambda_{-}\mid+\zeta\lambda_{+}}{\mid\lambda_{-}\mid
+\lambda_{+}})I_{9}+(\frac{\eta-\zeta}{\mid\lambda_{-}\mid
+\lambda_{+}})W_{c}.\ee In this case
$Q_{0}=\mid\psi_{00}><\psi_{00}\mid+\mid\psi_{10}><\psi_{10}\mid+\mid\psi_{20}><\psi_{20}\mid$
and  by substituting this result in the Eq.(\ref{nd5}) we get
$$\rho=(\mu-\frac{\eta-\zeta}{3(\mid\lambda_{-}\mid
+\lambda_{+})})\mid\psi_{00}><\psi_{00}\mid+(\mu+(12x-1)\frac{\eta-\zeta}{3(\mid\lambda_{-}\mid
+\lambda_{+})})\mid\psi_{10}><\psi_{10}\mid$$ $$
+(\mu-(12x-2)\frac{\eta-\zeta}{3(\mid\lambda_{-}\mid
+\lambda_{+})})\mid\psi_{20}><\psi_{20}\mid)+\frac{\eta(\mid
\lambda_{-}\mid-\frac{1}{6})-\zeta(\mid
\lambda_{+}\mid+\frac{1}{6})}{3(\mid\lambda_{-}\mid
+\lambda_{+})})(\mid\psi_{01}><\psi_{01}\mid
$$ \be+\mid\psi_{02}><\psi_{02}\mid+\mid\psi_{11}><\psi_{11}\mid+
\mid\psi_{22}><\psi_{22}\mid+\mid\psi_{12}><\psi_{12}\mid+\mid\psi_{21}><\psi_{21}\mid).\ee
The positivity of $\rho$ requires that  all the Bell states
diagonal operator coefficients to be positive, and that this
condition be imposed on the coefficient $\mu$ only. So we get
\be\label{mu}\left\{\begin{array}{cc} x\geq \frac{1}{8} & \mu \geq
\frac{(12x-1)(\frac{1}{3}-2\eta)}{(12x-1)+3(\mid\lambda_{-}\mid
+\lambda_{+})}\\ x\leq \frac{1}{8} & \mu \geq
\frac{(2-12x)(\frac{1}{3}-2\eta)}{(2-12x)+3(\mid\lambda_{-}\mid
+\lambda_{+})}, \end{array}\right.\ee which, in this case means
$Q_{0}$ is on the boundary. Now by using this bound entangled BD
state we can find n-d condition for BDEW. We know EW will be an
nd-EW  if this EW is able to detect any bound state. Then by
using the equations (\ref{nd6}) and (\ref{nd7}) we have \be
Tr(W\rho)=(\frac{\varepsilon\mu}{3}+3(\frac{\varepsilon}{9\lambda_{+}}+(1-\varepsilon))\eta\lambda_{+}+3(\frac{\varepsilon}{9\lambda_{-}}-(1-\varepsilon))\zeta\mid\lambda_{-}\mid<0.
\ee Now  by substituting $\varepsilon$ from Eq.(\ref{epsilon}) we
obtain \be
{\bf{r}}<\frac{-3+27(\zeta\mid\lambda_{-}\mid-\eta\lambda_{+})}{1+27(\zeta\mid\lambda_{-}\mid-\eta\lambda_{+})},
\ee where the calculated {\bf{r}} is greater than the represented
{\bf{r}} for EW in Eq.(\ref{nd4}). Therefore, we can find one of
the p's corresponding to EW which is an nd-EW. Non-decomposable
generalized EW for a general case is under investigation.
\section{Choi map} Choi positive map \cite{choi}
$\phi(a,b,c):M^{3}\rightarrow M^{3}$ is defined as
 \be \phi_{a,b,c}(\rho)=\left(\begin{array}{ccc}
a\rho_{11}+b\rho_{22}+c\rho_{33} & 0 & 0 \\0 &
a\rho_{22}+b\rho_{33}+c\rho_{11} & 0 \\ 0 & 0 &
a\rho_{33}+b\rho_{11}+c\rho_{22}\end{array}\right)-\rho, \ee
where  $\rho\in M^{3}$. It was shown that $\phi(a,b,c)$ is
positive iff \be a\geq 1\;\;,\;\; a+b+c\geq 3\;\;,\;\; 1\leq
a\leq 3. \ee Using  Jamiolkowski \cite{jamiolkowski} isomorphism
between the positive map  and the  operators we obtain the
following $3\otimes 3$ EW corresponding to Choi map
 \be
W_{Choi}=\frac{1}{3(a+b+c-1)}(a\sum_{k=0}^{2}\ket{\psi_{k0}}\bra{\psi_{k0}}+b\sum_{k=0}^{2}\ket{\psi_{k2}}\bra{\psi_{k2}}+c\sum_{k=0}^{2}\ket{\psi_{k1}}\bra{\psi_{k1}}-3\ket{\psi_{00}}\bra{\psi_{00}}).
\ee Similar to BDEW we expand $\ket{\psi_{00}}\bra{\psi_{00}}$
using the identity operator and the other Bell diagonal
states:\be \ket{\psi_{00}}\bra{\psi_{00}}=I_{9}-\sum_{i\ne
j=0}^{2}\ket{\psi_{ij}}\bra{\psi_{ij}}. \ee Then we reduce EW to
the following form $$
W_{Choi}=\frac{1}{3(a+b+c-1)}(-(3-a)I_{9}+3\sum_{k=1}^{2}\ket{\psi_{k0}}\bra{\psi_{k0}}$$
\be\label{choi4}
+(b+3-a)\sum_{k=0}^{2}\ket{\psi_{k2}}\bra{\psi_{k2}}+(c+3-a)\sum_{k=0}^{2}\ket{\psi_{k1}}\bra{\psi_{k1}}).\ee
Comparing with BDEW (\ref{GWE1}) we have \be
{\bf{r}}=-\frac{3(3-a)}{(a+b+c-1)}, \ee and the  EW operator is
defined as $$
W_{Choi}={\bf{r}}I_{9}/9+(1-{\bf{r}})(\frac{1}{(8-2a+b+c)}\sum_{k=1}^{2}\ket{\psi_{k0}}\bra{\psi_{k0}}+\frac{(b+3-a)}{3(8-2a+b+c)}\sum_{k=0}^{2}\ket{\psi_{k2}}\bra{\psi_{k2}}$$
\be\label{choi1}+\frac{(c+3-a)}{3(8-2a+b+c)}\sum_{k=0}^{2}\ket{\psi_{k1}}\bra{\psi_{k1}}).
\ee By comparing (\ref{choi4}) with (\ref{GWE1}) we obtain the
coefficients $q_{ij}$ $$
q_{10}=q_{20}=\frac{1}{(8-2a+b+c)}\;\;,\;\;q_{02}=q_{12}=q_{22}=\frac{b+3-a}{3(8-2a+b+c)},$$
\be\label{choi2} q_{01}=q_{11}=q_{21}=\frac{c+3-a}{3(8-2a+b+c)}.
\ee Note that if $r$ is negative, as  introduced in EW above, this
operator will be  positive, but not a completely positive map. For
${\bf{r}}\leq 0$  we have $1 \leq a\leq 3$. By  assuming $a\geq
b\geq c$, the  minimum negative eigenvalue of choi EW
(\ref{choi1}) is given by \be
\frac{{\bf{r}}}{9}+(1-{\bf{r}})\frac{c+3-a}{3(8-2a+b+c)}< 0,\ee
where upon  substituting {\bf{r}} from  Eq.(\ref{choi2}) we get
$1 \leq a\leq 2$. This is equal to the introduced positivity
condition of Choi map in \cite{choi}.

By using (\ref{optt1}) for non-negativity of the observable
$W_{choi}$ we find the distributions $P_{ij}$ as a function of
$q_{ij}$. The minimum value of $C_{\gamma}$  is obtained  from
the boundary of the feasible region, i.e., we have \be
(C_{\gamma})=\frac{1}{(8-2a+b+c)}{\cal
P}_{1}+\frac{(b+3-a)}{3(8-2a+b+c)}{\cal
P}_{2}+\frac{(c+3-a)}{3(8-2a+b+c)}{\cal P}_{3}, \ee where ${\cal
P}_{1}=\sum_{k=1}^{2}P_{_{k0}},{\cal
P}_{2}=\sum_{k=0}^{2}P_{_{k2}}$ and ${\cal
P}_{3}=\sum_{k=0}^{2}P_{_{k1}}$. We can find the extreme value of
$({\cal P}_{1},{\cal P}_{2},{\cal P}_{3})$ which is  obtained
under the product states
$\ket{\gamma}=\ket{\alpha}_{1}\ket{\alpha}_{2}$ as
 \be\left\{\begin{array}{c} {\cal P}_{1}=\mid\alpha_{1}\mid^2\mid\beta_{1}\mid^2+\mid\alpha_{2}\mid^2\mid\beta_{2}\mid^2+\mid\alpha_{3}\mid^2\mid\beta_{3}\mid^2\\ -\frac{1}{3}\mid \mid\alpha_{1}\mid
\mid\beta_{1}\mid+\mid\alpha_{2}\mid \mid\beta_{2}\mid
e^{i\phi_{2}}+\mid\alpha_{3}\mid
\mid\beta_{3}\mid e^{i\phi_{3}}\mid ^2\\
{\cal
P}_{2}=\mid\alpha_{1}\mid^2\mid\beta_{1}\mid^2+\mid\alpha_{3}\mid^2\mid\beta_{1}\mid^2+\mid\alpha_{3}\mid^2\mid\beta_{2}\mid^2\\{\cal
P}_{3}=\mid\alpha_{1}\mid^2\mid\beta_{2}\mid^2+\mid\alpha_{2}\mid^2\mid\beta_{3}\mid^2+\mid\alpha_{3}\mid^2\mid\beta_{1}\mid^2\end{array}\right.,
\ee where $\ket{\alpha}_{1}=\left(\begin{array}{c} \alpha_{1}\\
\alpha_{2}\\ \alpha_{3}\end{array}\right)$ and $\ket{\alpha}_{1}=\left(\begin{array}{c} \beta_{1}\\
\beta_{2}\\ \beta_{3}\end{array}\right)$. One can obtain the
extreme points of the $({\cal P}_{1},{\cal P}_{2},{\cal P}_{3})$
as \be \left\{\begin{array}{ccc}
\ket{\alpha}_{1}=\ket{\alpha}_{2}=\left(\begin{array}{c}1\\ \omega
\\ \bar{\omega}\end{array}\right) & \rightarrow &
({\cal P}_{1}=\frac{1}{3},{\cal P}_{2}=\frac{1}{3},{\cal
P}_{3}=\frac{1}{3})
\\ \ket{\alpha}_{1}=\left(\begin{array}{c}1\\0\\
0\end{array}\right),
\ket{\alpha}_{2}=\left(\begin{array}{c}0\\0\\
1\end{array}\right) & \rightarrow & ({\cal P}_{1}=0,{\cal
P}_{2}={1},{\cal P}_{3}=0)
\\ \ket{\alpha}_{1}=\left(\begin{array}{c}1\\ 0\\ 0\end{array}\right), \ket{\alpha}_{2}=\left(\begin{array}{c}0\\ 1\ 0\end{array}\right)
& \rightarrow & ({\cal P}_{1}=0,{\cal
P}_{2}=0,{\cal P}_{3}=1) \\
\ket{\alpha}_{1}=\left(\begin{array}{c}1\\
0\\ 0\end{array}\right), \ket{\alpha}_{2}=\left(\begin{array}{c}1\\
0\\ 0 \end{array}\right) & \rightarrow & ({\cal
P}_{1}=\frac{2}{3},{\cal P}_{2}=0,{\cal
P}_{3}=0)\end{array}\right. .\ee The convex combination of all
extreme points provide a convex or a feasible region (Fig-3), then
we have the following optimization problem \be
\left\{\begin{array}{cc} \mbox{minimize}&
(C_{\gamma})=(\frac{1}{(8-2a+b+c)}{\cal
P}_{1}+\frac{(b+3-a)}{3(8-2a+b+c)}{\cal
P}_{2}+\frac{(c+3-a)}{3(8-2a+b+c)}{\cal P}_{3}\\
\mbox{subject to} & 1-\frac{3}{2}{\cal P}_{1}-{\cal
P}_{2}-\frac{1}{2}{\cal P}_{3}\leq 0 \\& 1-\frac{3}{2}{\cal
P}_{1}-\frac{1}{2}{\cal P}_{2}-{\cal P}_{3}\leq 0\\& 1-{\cal
P}_{1}-{\cal P}_{2}-{\cal P}_{3}\geq 0 \\ & {\cal P}_{1},{\cal
P}_{2},{\cal P}_{3}\geq 0.\end{array}\right. \ee Whether
analytically  we have been able to show that we will have
violation only from the two planes
$$2-{3}{\cal P}_{1}-2{\cal
P}_{2}-{\cal P}_{3}=0$$ $$2-{3}{\cal P}_{1}-2{\cal P}_{3}-{\cal
P}_{2}=0.$$ Now let us  assume that the maximum value of  the
violation from the planes is $\Delta<1$. Thus, the equation of
the plane passing through the new extreme points, parallel to the
above plane, is obtained. Next we derive the intersection of the
following adjacent  planes \be\left\{\begin{array}{cc} 1)& 3{\cal
P}_{1}+{\cal P}_{2}+2{\cal P}_{3}-(2+\Delta)=0 \\ 2) & 3{\cal
P}_{1}+2{\cal P}_{2}+{\cal P}_{3}-(2+\Delta)=0 \\ 3) &  {\cal
P}_{1}+{\cal
P}_{2}+{\cal P}_{3}-1=0 \\ 4) &  {\cal P}_{1}=0 \\ 5) & {\cal P}_{2}=0 \\ 6) &  {\cal P}_{3}=0 \\
7) &  {\cal P}_{1}=\frac{2}{3} \\ 8) &   {\cal P}_{2}=1 \\
9) &  {\cal P}_{3}=1 \end{array}\right. ,\ee where  new extreme
points are obtained from intersecting the above planes. Next we
calculate $C_{\gamma}$ for all the newly obtained extreme points
and compare them with each other. Some easy calculations gives the
minimum value of the parameter $C_{\gamma}$ which is independent
from $\Delta$  \be
(C_{\gamma})_{{min}}=\frac{6+2(c-a)}{9(8-2a+b+c)},\ee then the
critical value of the  parameter r is obtained as \be {\bf
r}_{c}=\frac{-6+2(a-c)}{2+b-c}. \ee For  $a=b=c=1$ the parameter r
reduces to $r_{c}=-3$  corresponding to the well known reduction
map. On the other hand,  EW (\ref{choi1}) must have positive trace
under any product state $\ket{\gamma}\bra{\gamma}$. Thus the
introduced r in EW must satisfy \be r\geq r_{c}\Rightarrow
\frac{-3(3-a)}{(a+b+c-1)}\geq \frac{-6+2(a-c)}{2+b-c}, \ee where
the inequality  is satisfied for all value of $0\leq a\leq 2$ and
$a\geq b\geq c$.
\section{Some separable states at the boundary of separable region}
Here we introduce some set of separable states as
$$\rho_{m}=\sum_{_{k}}\ket{\psi_{km}}\bra{\psi_{km}}=\sum_{_{l}}\ket{l}\bra{l}\otimes\ket{l+m}\bra{l+m},$$
$$\rho^{\prime}_{m}=\sum_{_{k}}\ket{\psi_{mk}}\bra{\psi_{mk}}=\sum_{_{l,l^{\prime},k}}\omega^{m(l-l^{\prime})}\ket{l}\bra{l^{\prime}}\otimes\ket{l+k}\bra{l^{\prime}+k},$$
\be\label{product}\rho^{\prime\prime}_{n}=\sum_{_{k}}\ket{\psi_{nk,k}}\bra{\psi_{nk,k}}=\sum_{_{l,l^{\prime},k}}\omega^{nk(l-l^{\prime})}\ket{l}\bra{l^{\prime}}\otimes\ket{l+k}\bra{l^{\prime}+k},
\ee where $n=0,1,2\;\;,\;\;m=0,1,2$. One can show that the  convex
sum of $\rho_{0}\;,\;\rho^{\prime\prime}_{0}=\rho^{\prime}_{0}$,
i.e, $\rho_{\mu}^{S}=\mu\rho_{0}+(1-\mu)\rho^{\prime}_{0},$ is
orthogonal to the optimal $W_{\Lambda}=\Lambda
W_{c}+(1-\Lambda)W_{red},$ i.e., we have
$Tr(W_{\Lambda}\rho_{S}^{\mu})=0$.

Hence, $\rho_{\mu}^{S}$ lie at the boundary of the separable
region \cite{horodecki4}. On the other hand, one can show that by
acting the  local unitary operation $U_{ij}$ over $W_{\Lambda}$ as
$(W_{\Lambda})_{ij}=U_{ij}(W_{\Lambda})U_{ij}^{\dagger}$ he
obtains a new set of optimal EW, $(W_{\Lambda})_{ij}$, the
application of which is not only to get a new set of bound
entangled states by acting local unitary operation, but also to
obtain some separable states
${(\rho^{\mu}_{S})}_{ij}=U_{ij}\rho^{\mu}_{S}U_{ij}^{\dagger}$ as
such  which are the convex sum of separable states (\ref{product})
at the boundary of separable states.
\section{Conclusion}
We have shown that finding  generic  Bell states diagonal
entanglement witnesses (BDEW) for $d_{1}\otimes d_{2}\otimes
....\otimes d_{n}$ systems has reduced to a linear programming
problem. Since solving linear programming for generic case is
difficult we have considered the following special cases. Also we
have considered BDEW for multi-qubit, $2\otimes N$ and $3\otimes
3$ systems and then have  considered optimality condition for $3
\otimes 3$ EW. Also, we have considered an n-d condition over $3
\otimes 3$ BDEW and have obtained this condition for some special
cases exactly. We have defined extensive group of nd-BDEW by
combining critical EW and the reduction map (each with special
coefficients). Then we have defined the Bell decomposable bound
entangled state and have considered detection of this state with
optimal BDEW and a general BDEW. Finally, we have considered Choi
map as an example of BDEW. Optimality and non-decomposibility of
EW for multi-qubit and $2\otimes N$ as well as  EW for generic
bipartite $d_{1}\otimes d_{2}$ systems and multipartite
$d_{1}\otimes d_{2}\otimes ...\otimes d_{n}$ are under
investigation.  As a physical implementation of EW we know that
the  optimization of decomposition of EW to find  the smallest
number of measurements possible for local measurement on a system
can be used. Therefore to make use of this implementation of EW
for the obtained EW's  is currently under investigation.

 \vspace{1cm}\setcounter{section}{0}
\setcounter{equation}{0}
\renewcommand{\theequation}{A-\roman{equation}}
{\Large APPENDIX A}

{\bf Minimization of the product distributions:}

In Eq.(3-5) the Bell orthonormal states for a $d_{1}\otimes
d_{2}\otimes ...\otimes d_{n}$ ($d_{1}\leq d_{2}\leq ...\leq
d_{n}$) have been introduced  by applying local unitary operation
on $\ket{\psi_{_{00}}}$. Let us further consider  a pure product
state
$\ket{\gamma}=\ket{\alpha}_{1}\ket{\alpha}_{2}...\ket{\alpha}_{n}$.
Then the product distributions can be written as \be
P_{_{i_{1},i_{2},...,i_{n}}}(\gamma)=\mid
<\gamma\mid\psi_{_{i_{1},i_{2},...,i_{n}}}>\mid^{2}. \ee It
easily follows that \be 0\leq
P_{_{i_{1},i_{2},...,i_{n}}}(\gamma)\leq \frac{1}{d_{1}}. \ee On
the other hand, from the completeness of Bell states:\be
\sum_{_{i_{1},i_{2},...,i_{n}}}\ket{\psi_{_{i_{1},i_{2},...,i_{n}}}}\bra{\psi_{_{i_{1},i_{2},...,i_{n}}}}=I_{d_{1}}\otimes
I_{d_{2}}\otimes ... \otimes I_{d_{n}}, \ee we have
$\sum_{_{i_{1},i_{2},...,i_{n}}}P_{_{i_{1},i_{2},...,i_{n}}}(\gamma)=1$,
which leads to \be \sum_{_{i_{1},i_{2},...,i_{n}}}\mid
<\gamma\mid\psi_{_{i_{1},i_{2},...,i_{n}}}>\mid^{2}=d_{1}. \ee The
above equation indicates that if we can show that for a
particular choice of $\ket{\alpha}_{i}$'s, the $d_{1}$-number  of
$\mid
<\gamma\mid\psi_{_{i_{1},i_{2},...,i_{n}}}>\mid^{2}=P_{_{i_{1},i_{2},...,i_{n}}}$
can have their  maximum value equal to $\frac{1}{d_{1}}$, then the
remaining ones will be zero.

To minimize the summation $C=\sum_{ij}q_{ij}P_{ij}$ for a $3
\otimes 3$ system, assuming  that $q_{00}=0$, let us first suppose
that $\ket{\alpha}=\ket{\beta}$ so that $P_{_{00}}=\frac{1}{3}$.
Then we  find the set
$\mid\bra{\alpha}U_{ij}\ket{\beta}\mid^{2}=1$ for different
possible choices of $\ket{\alpha}$ and $U_{ij}$:
$$
\ket{\alpha}=\begin{array}{cccccccc}\left(\begin{array}{c} 1 \\ 1
\\1
\end{array}\right) & ,& \left(\begin{array}{c} 1 \\ \omega \\ \bar{\omega}
\end{array}\right) & ,& \left(\begin{array}{c} 1 \\ \bar{\omega} \\ {\omega}
\end{array}\right) &, &
\ket{\psi_{01}},\ket{\psi_{02}}, &
\mbox{min}(\sum_{ij}q_{ij})=q_{_{01}}+q_{_{02}},\end{array}$$
$$ \ket{\alpha}=\begin{array}{cccccccc}\left(\begin{array}{c} 1 \\
0
\\0
\end{array}\right) & , & \left(\begin{array}{c} 0 \\ 1 \\
0
\end{array}\right) & , & \left(\begin{array}{c} 0 \\ 0 \\
1
\end{array}\right) &,&
\ket{\psi_{10}},\ket{\psi_{20}}, &
\mbox{min}(\sum_{ij}q_{ij})=q_{_{10}}+q_{_{20}},\end{array}$$
$$
\ket{\alpha}=\begin{array}{cccccccc}\left(\begin{array}{c} 1 \\ 1
\\ \omega
\end{array}\right) & , & \left(\begin{array}{c} 1 \\ \omega \\
1
\end{array}\right) & , & \left(\begin{array}{c} \omega \\ 1 \\
1
\end{array}\right) &,&
\ket{\psi_{11}},\ket{\psi_{22}}, &
\mbox{min}(\sum_{ij}q_{ij})=q_{_{11}}+q_{_{22}},\end{array}$$

$$
\ket{\alpha}=\begin{array}{cccccccc}\left(\begin{array}{c} 1 \\ 1
\\ \bar{\omega}
\end{array}\right) & , & \left(\begin{array}{c} 1 \\ \bar{\omega} \\
1
\end{array}\right) & , & \left(\begin{array}{c} \bar{\omega} \\ 1 \\
1
\end{array}\right) &,&
\ket{\psi_{12}},\ket{\psi_{21}}, &
\mbox{min}(\sum_{ij}q_{ij})=q_{_{12}}+q_{_{21}}.\end{array}$$ The
above relations imply that  $C_{mn}=\frac{1}{3}(q_{1}+q_{2})$,
where $q_{1}$ and $q_{2}$ correspond to two of $q_{ij}$ appearing
in the same  row.

\vspace{1cm} \setcounter{section}{0} \setcounter{equation}{0}
\renewcommand{\theequation}{B-\roman{equation}}

{\Large APPENDIX B}

{\bf Critical entanglement witness is optimal:}

According to the References \cite{cirac,kraus}, an  EW will be
optimal if for all positive operator P and $\varepsilon>0$, the
operator \be\label{appp2}
W^{\prime}=(1+\varepsilon)W_{c}-\varepsilon P \ee is not an EW. In
order to prove  the critical EW given in (\ref{op8}) is optimal,
we first show that \be\label{appp1}
Tr(W_{c}\ket{\alpha}\bra{\alpha}\otimes
\ket{\alpha^{\ast}}\bra{\alpha^{\ast}})=0. \ee It just suffices
to check that for the product distribution $
P_{ij}=<\psi_{ij}\ket{\alpha}\bra{\alpha}\otimes
\ket{\alpha^{\ast}}\bra{\alpha^{\ast}}\psi_{ij}>$,  we have $
P_{00}=\frac{1}{3}\;\;,\;\;P_{01}=P_{02}\;\;,\;\;P_{11}=P_{22}\;\;,\;\;P_{12}=P_{21}$.

Substituting $P_{ij}$ given above in (\ref{appp1}), it is easy to
see that $Tr(W_{c}\ket{\alpha}\bra{\alpha}\otimes
\ket{\alpha^{\ast}}\bra{\alpha^{\ast}})=0$. Also it is
straightforward to see that there exists no positive operator P
with the constraint$$ Tr(P\ket{\alpha}\bra{\alpha}\otimes
\ket{\alpha^{\ast}}\bra{\alpha^{\ast}})=0\;\;,\;\;\forall
\ket{\alpha}$$. Therefore, there exist  no positive operator
{\bf{r}} to satisfy (\ref{appp2}). Hence $W_{c}$, and in
particular $W_{red}$, are optimal.

 \vspace{1cm} \setcounter{section}{0}
\setcounter{equation}{0}
\renewcommand{\theequation}{B-\roman{equation}}

{\Large APPENDIX C}

{\bf Simplex method for solving multi-qubit minimization problem}

We know that simplex method  is an elegant way for solving linear
programming problems. As an example we obtain the $P_{_{00...00}}$
and $P_{_{10...00}}$ constraints in  Eq.(\ref{multi1}), thus we
have two slack variables which are defined as \be
\omega_{1}=\frac{1}{2^{n-1}}-
2P_{_{00...00}}+2P_{_{10...00}}(1-\frac{1}{2^{n-1}})\;\;,\;\;\omega_{2}=\frac{1}{2^{n-1}}-
2P_{_{10...00}}+2P_{_{00...00}}(1-\frac{1}{2^{n-1}}).\ee We carry
out this procedure to transform the inequality constraints (4-31)
into  equality  \be\label{appc} \left\{\begin{array}{cc}
\mbox{maximize}&
-C_{\gamma}=\frac{1}{2(2^{n-1}-1)}(-(1-x)+(1-x)P_{_{00...00}}+((2^{n}-1)x-1)P_{_{10...00}})\\
\mbox{subject to} & \omega_{1}=\frac{1}{2^{n-1}}-
2P_{_{00...00}}+2P_{_{10...00}}(1-\frac{1}{2^{n-1}})\\
& \omega_{2}=\frac{1}{2^{n-1}}-
2P_{_{10...00}}+2P_{_{00...00}}(1-\frac{1}{2^{n-1}})\\ &
P_{_{00...00}},P_{_{10...00}},\omega_{1},\omega_{2}\geq
0.\end{array}\right. \ee Now we rewrite the first equation in
(B-ii) in terms of $\omega_{1}$ and $\omega_{2}$, making use of
the slack variables: \be
-C_{\gamma}=\frac{1}{2(2^{n-1}-1)}(-(1-x)+\frac{(1-x)a-(2^{n}-1)x+1}{2(a^{2}-1)}\omega_{1}+\frac{(1-x)-((2^{n}-1)x-1)a}{2(a^{2}-1)}\omega_{2}
,\ee where $a=1-\frac{1}{2^{n-2}}$. For $0\leq x\leq
\frac{1}{2^{n-1}+1}$  the coefficients $\omega_{1}$ and
$\omega_{2}$ are both negative. Now  from the simplex method we
conclude  $\omega_{1}=\omega_{2}=0$ , i.e.,
$P_{_{00...00}}=P_{_{10...00}}=\frac{1}{2}$. Thus the minimum
value of $C_{\gamma}=\frac{x}{2}$. For $ \frac{1}{2^{n-1}+1}\leq
x\leq 1$, from (\ref{appc}), we see that the coefficient
$P_{_{10...00}}$ is negative, so that $P_{_{10...00}}=0$, hence
$P_{_{00...00}}=\frac{1}{2^{n-1}}$. Therefore, we  find the
minimum value of $C_{\gamma}$ as
$(C_{\gamma})_{min}=\frac{1-x}{2^{n}}$.

\newpage
{\bf Figure Captions}

{\bf Figure-1:} Feasible region for $3\otimes 3$ systems for
particular choice of $q_{_{00}}=0$, $q_{_{10}}=x$,
$q_{_{20}}=\frac{1}{4}-x$ and others $q$'s being equal, i.e.,
when the linear programming variables are $P_{_{00}}$,
$P_{_{10}}$ and $P_{_{20}}$ .

{\bf Figure-2:} Feasible region for $3\otimes 3$ systems for
particular choice of $q_{_{00}}=0$, $q_{_{10}}=x$,
$q_{_{01}}=\frac{1}{4}-x$ and others $q$'s being equal, i.e.,
when the linear programming variables are $P_{_{00}}$,
$P_{_{10}}$ and $P_{_{01}}$ .

{\bf Figure-3:} Feasible region for $3\otimes 3$  Choi map for
particular choice of $a \geq b\geq c$, i.e., when the linear
programming variables are ${\cal P}_{1}\sum_{k=1}^{2}P_{_{k0}},
{\cal P}_{2}\sum_{k=0}^{2}P_{_{k2}}$ and  ${\cal
P}_{1}\sum_{k=0}^{2}P_{_{k1}}$.


\begin{thebibliography}{99}
\bibitem{Einstein} A. Einstein, B. Podolsky, and N. Rosen, Phys. Rev. {\bf 47}, 777
(1935).
\bibitem{nielsen} M. N. Nielsen and I. L. Chuang, Quantum
computation and quantum information (Cambridge University Press,
Cambridge, 2000).
\bibitem{werner} R. F. Werner, Phys. Rev. A {\bf 40},
4277 (1989).
\bibitem{peres} A. Peres, Phys. Rev. Lett. {\bf 77}, 1413
(1996).
\bibitem{horodecki1} M. Horodecki, P. Horodecki, and R. Horodecki, Phys. Lett. A {\bf
223}, 1 (1996).
\bibitem{woronowicz} S. L. Woronowicz, Rep. on Math. Phys. {\bf  10}, 165
(1976).
\bibitem{horodecki2} P. Horodecki, Phys. Lett. A {\bf  232}, 333
(1997).
\bibitem{horodecki3} M. Horodecki, P. Horodecki, and R. Horodecki, Phys.
Rev. Lett. {\bf 80}, 5239 (1998).
\bibitem{terhal} B. M. Terhal, Phys. Lett. A{\bf  271}, 319
(2000).
\bibitem{jamiolkowski} A. Jamiolkowski, Rep. Mat, Phys,
{\bf 3}, 275 (1972).
\bibitem{terhal2} B. M. Terhal, Journal of
Theoretical Computing Science {\bf 287}(1), 313 (2002).
\bibitem{bruss} D. Bruss et al., J. Mod. Opt. {\bf 49}, 1399 (2002).
\bibitem{Doherty} A. C. Doherty,P. A. Parrilo and F. M. Spedalieri, Phys. Rev. A{\bf 69}, 022308
(2004).
\bibitem{cirac} M. Lewenstein, B. Kraus, J. I. Cirac, P.
Horodecki, Phys. Rev. A{\bf  62}, 052310 (2000).
\bibitem{kraus}M. Lewenstein, B. Kraus, P. Horodecki, and J. I. Cirac, Phys.
Rev. A{\bf  63}, 044304 (2001).
\bibitem{choi} M. D. Choi, Linear Algebra and its Applications {\bf 12}, 95 (1975).
\bibitem{lec} M. Lewenstein, Quantum Information Theory, Institute
for Theoretical Physics, Unversity of Hannover, March 31, (2004).
\bibitem{Otfrid1} O. G\"{u}hne and P. Hyllus, Int. J. Theor. Phys. {\bf 42}, 1001 (2003).
\bibitem{Otfrid2} G. T\'{o}th and O. G\"{u}hne, Phys. Rev. Lett {\bf  94}, 060501 (2005).
\bibitem{Otfrid3} G. T\'{o}th and O. G\"{u}hne, AIP. Conf. Proc. {\bf 734}, 234 (2004).
\bibitem{Hyllus} O. G\"{u}hne, P. Hyllus and D. Bruss, J. Mod. Opt. {\bf 50} (6-7), 1079 (2003).
\bibitem{Rob1} G. Vidal, R.Tarrach, Phys. Rev. A{\bf  59}, 141 (1999).
\bibitem{Rob2} J. F. Du, M. J. Shi, X. Y. Zhou, R. D. Han, Phys. Lett. A{\bf  264}, 244 (2000).
\bibitem{boyd} S. Boyd and L. Vandenberghe,
\emph{Convex Optimization}, Cambridge University Press, (2004).
\bibitem{lsd} M. Lewenstein
and A. Sanpera, Phys. Rev. Lett. {\bf 80}, 2261 (1998).
\bibitem{lsd1} M. A. Jafarizadeh, M. Mirzaee, M. Rezaee, Physica A.
{\bf 349}, 459, (2005).
\bibitem{lsd2} M. A. Jafarizadeh, M. Mirzaee, M. Rezaee,
International Journal of Quantum Information. {\bf 2}, 4, 541,
(2004).
\bibitem{horodecki4} R. Horodecki and M. Horodecki, Phys. Rev. A {\bf 54}, 1838 (1996).
\end{thebibliography}
\end{document}